\documentclass[11pt,twoside]{article}
\input{epsf.sty}
\usepackage{mathrsfs}
\usepackage{amsmath}
\usepackage{amssymb}
\usepackage{cite}
\usepackage{appendix}
\usepackage{graphicx}
\usepackage{subfigure}
\newtheorem{theorem}{Theorem}
\newtheorem{definition}{Definition}
\newtheorem{proposition}{Proposition}

\numberwithin{equation}{section}
\newcommand*{\QEDB}{\hfill\ensuremath{\square}}

\title{The massive Thirring system in the quarter plane}
\author{Baoqiang Xia
\\
School of Mathematics and Statistics, Jiangsu Normal
University,\\
 Xuzhou, Jiangsu 221116, P. R. China,
 \\
 E-mail address:
xiabaoqiang@126.com
}

\date{}
\begin{document}
\maketitle
\begin{abstract}

The unified transform method (UTM) for analyzing initial-boundary value (IBV) problems
provides an important generalization of the inverse scattering transform (IST) method for analyzing initial value problems.
In comparison with the IST, a major difficulty of the implementation of the UTM in general is the involvement of unknown boundary values.
In this paper we analyze the IBV problem for the massive Thirring model posed in the quarter plane.
We show for this integrable model, the UTM is as effective as the IST method:
the Riemann-Hilbert (RH) problems we formulated for such a problem have explicit $(x, t)$-dependence and depend only on the given initial and boundary values;
they do {\it not} involve additional unknown boundary values.\\


\end{abstract}

\section{ Introduction}

The massive Thirring (MT) model, as an important solvable quantum field theory model, was introduced by Thirring in 1958 \cite{Thirr}.
In the laboratory coordinates, this model can be written in the form
\begin{eqnarray}
\begin{split}
i(u_t+u_x)+v+|v|^2u=0,
\\
i(v_t-v_x)+u+|u|^2v=0.
\end{split}
\label{MT}
\end{eqnarray}
It was established in \cite{Mik,KM} that (\ref{MT}) is integrable:
it is the compatibility condition of the following pair of linear spectral problems (called Lax pair)
\begin{subequations}
\begin{eqnarray}
\varphi_x-\frac{i}{4}(\lambda^2-\lambda^{-2})[\sigma_3,\varphi]=
Q\varphi,
\label{LPx1}
\\
\varphi_t-\frac{i}{4}(\lambda^2+\lambda^{-2})[\sigma_3,\varphi]=
P\varphi,
\label{LPt1}
\end{eqnarray}
\label{LP1}
\end{subequations}
where
\begin{eqnarray}
\begin{split}
\sigma_3=\left( \begin{array}{cc} 1 & 0 \\
 0 &  -1 \\ \end{array} \right),
\\
Q=\frac{i}{4}\left(|u|^2-|v|^2\right)\sigma_3-\frac{i\lambda}{2}\left( \begin{array}{cc} 0 & \bar{v} \\
v &  0 \\ \end{array} \right)+\frac{i}{2\lambda}\left( \begin{array}{cc} 0 & \bar{u} \\
u &  0 \\ \end{array} \right),
 \\
P=-\frac{i}{4}\left(|u|^2+|v|^2\right)\sigma_3-\frac{i\lambda}{2}\left( \begin{array}{cc} 0 & \bar{v} \\
v &  0 \\ \end{array} \right)-\frac{i}{2\lambda}\left( \begin{array}{cc} 0 & \bar{u} \\
u &  0 \\ \end{array} \right).
\end{split}
\label{QP}
\end{eqnarray}
The MT system (\ref{MT}) has been studied by numerous authors from different points of view
(for example, see \cite{KM,Orf,KMIST,VilIST,KLIST,PS1,KN1,Alo,Lee,NCQL,Pri,WS,EGH,NC,JP,Saa} and references therein).
In particular, the inverse scattering transform (IST) for the initial value problem of the MT system
was presented in \cite{KM,Orf,KMIST,VilIST,KLIST} and was recently studied in \cite{PS1}.

The purpose of this paper is to study the initial-boundary value (IBV) problem for the MT system (\ref{MT}) posed in the quarter plane
$\left\{(x,t)\in \mathbb{R}^2\mid~0\leq x<\infty,0\leq t<\infty\right\}$.
We assume the initial and Dirichlet boundary conditions are prescribed:
\begin{eqnarray}
\begin{split}
u(x,0)=u_0(x), ~~v(x,0)=v_0(x), ~~0\leq x<\infty,
\\
u(0,t)=g_0(t), ~~v(0,t)=h_0(t), ~~0\leq t<\infty,
\end{split}
\label{IBV}
\end{eqnarray}
where $u_0(x)$, $v_0(x)$, $g_0(t)$ and $h_0(t)$ are given functions of Schwartz class,
and they are compatible at $x=t=0$, i.e. $u_0(0)=g_0(0)$ and $v_0(0)=h_0(0)$.

The method we employ in the present paper is the unified transform method (UTM),
which was presented by Fokas \cite{F1,F6}.
This method provides an important generalization of the IST formalism from initial value problems to IBV problems.
The UTM has been used extensively in the literature for analyzing IBV problems for integrable nonlinear evolution equations;
see for example \cite{F3,F4,F5,MFS1,FL1,L1,L2,XF1,XF2,GLZ,Tian,BH1,BH3,XFPD,Xia} and references therein.
The implementation of this method to the IBV problem of an integrable PDE yields the solution in
terms of the solution of a matrix Riemann-Hilbert (RH) problem,
which is formulated via certain functions called spectral functions.
In comparison with the IST, a major difficulty of this problem in general is that the corresponding RH problem involves unknown boundary values.
For example, when the UTM is applied to the nonlinear Schr\"{o}dinger (NLS) equation in the quarter plane,
one needs both the Dirichlet and the Neumann boundary values to define the corresponding RH problem.
Thus one needs to characterize the unknown boundary values through the given initial and boundary data.
However, for the IBV problem of the MT system, we find that the UTM is as effective as the IST:
by performing the simultaneous spectral analysis of the associated Lax pair,
we can express the solution of the IBV problem of the MT system through the solution of a $2 \times 2$ matrix RH problem.
This RH problem has explicit $(x, t)$-dependence and is defined only in terms of the given initial data $(u_0(x),v_0(x))$ and Dirichlet boundary values $(g_0(t),h_0(t))$;
this RH problem does {\em not} involve additional unknown boundary values.

We note that the Lax pair (\ref{LP1}) involves rational dependence on the spectral parameter $\lambda$.
For convenience of performing spectral analysis, we transform both parts of (\ref{LP1}) to two equivalent forms.
Thus we derive two RH problems for the IBV problem of the MT system:
one problem is convenient for recovering the component $u$ as $\lambda\rightarrow 0$,
whereas the other problem is convenient for recovering the component $v$ as $\lambda\rightarrow \infty$.
This idea is borrowed from papers \cite{VilIST,PS1} in which a similar strategy was used to study the IST for the MT system.

The main results of this paper are stated in Theorem 1 and Theorem 2.
The plan of this paper is as follows.
In section 2, we transform the Lax pair (\ref{LP1}) to two equivalent forms
which are convenient for performing the spectral analysis.
In section 3, we consider the direct part of this problem: we assume a solution exists.
By performing the spectral analysis of the two Lax pairs presented in section 2,
we formulate two RH problems and we express the solution of the problem in terms of the solutions of these two RH problems.
In section 4, we study the inverse part of the problem:
we prove that the expressions for the solution constructed in section 3
do indeed satisfy the MT system as well as the given initial-boundary conditions.
We discuss further our results in section 5.
A few details on the derivations of some results are presented in appendices.

\section{Transformations of the Lax pair}

Following \cite{PS1}, we transform the Lax pair (\ref{LP1}) with the spectral parameter $\lambda$ to two equivalent forms:
one, with the new spectral parameter $k$, is suitable to study the behaviour for $\lambda $ near origin,
and the other one, with the new spectral parameter $z$, is suitable to study the behaviour for $\lambda$ near infinity.
More precisely, we introduce the following two transformations
\begin{eqnarray}
\Phi=T(\lambda;u)\varphi,
~~
T(\lambda;u)=\left( \begin{array}{cc} 1 & 0 \\
 u &  \lambda^{-1} \\ \end{array} \right),
 \label{T1}
 \\
 \Psi=\mathcal{T}(\lambda;u)\varphi,
 ~~
 \mathcal{T}(\lambda;u)=\left( \begin{array}{cc} 1 & 0 \\
 v &  \lambda \\ \end{array} \right),
 \label{T2}
\end{eqnarray}
and we introduce the new spectral parameters
\begin{eqnarray}
k\triangleq\lambda^{-2},~~z\triangleq\lambda^2.
\label{newsp}
\end{eqnarray}
After straightforward calculations, we find that the transformation (\ref{T1}) maps the Lax pair (\ref{LP1}) to
\begin{subequations}
\begin{eqnarray}
\Phi_x-\frac{i}{4}(k^{-1}-k)[\sigma_3,\Phi]=
U\Phi,
\label{LPkx}
\\
\Phi_t-\frac{i}{4}(k^{-1}+k)[\sigma_3,\Phi]=
V\Phi,
\label{LPkt}
\end{eqnarray}
\label{LPk}
\end{subequations}
with
\begin{eqnarray}
\begin{split}
U=\left( \begin{array}{cc} -\frac{i}{4}\left(|u|^2+|v|^2\right) & \frac{i}{2}\bar{u} \\
u_x-\frac{i}{2}u|v|^2-\frac{i}{2}v &  \frac{i}{4}\left(|u|^2+|v|^2\right) \\ \end{array} \right)
+\frac{i}{2k}\left( \begin{array}{cc} u \bar{v} & -\bar{v} \\
u+u^2\bar{v} &  -u \bar{v} \\ \end{array} \right)\triangleq U_1+\frac{i}{2k}U_2,
 \\
V=\left( \begin{array}{cc} \frac{i}{4}\left(|u|^2-|v|^2\right) & -\frac{i}{2}\bar{u} \\
u_t-\frac{i}{2}u|v|^2-\frac{i}{2}v &  -\frac{i}{4}\left(|u|^2-|v|^2\right) \\ \end{array} \right)
+\frac{i}{2k}\left( \begin{array}{cc} u \bar{v} & -\bar{v} \\
u+u^2\bar{v} &  -u \bar{v} \\ \end{array} \right)\triangleq V_1+\frac{i}{2k}U_2;
\end{split}
\label{UV}
\end{eqnarray}
whereas the transformation (\ref{T2}) maps the Lax pair (\ref{LP1}) to
\begin{subequations}
\begin{eqnarray}
\Psi_x-\frac{i}{4}(z-z^{-1})[\sigma_3,\Psi]=
\mathcal{U}\Psi,
\label{LPzx}
\\
\Psi_t-\frac{i}{4}(z+z^{-1})[\sigma_3,\Psi]=
\mathcal{V}\Psi,
\label{LPzt}
\end{eqnarray}
\label{LPz}
\end{subequations}
with
\begin{eqnarray}
\begin{split}
\mathcal{U}=\left( \begin{array}{cc} \frac{i}{4}\left(|u|^2+|v|^2\right) & -\frac{i}{2}\bar{v} \\
v_x+\frac{i}{2}|u|^2v+\frac{i}{2}u &  -\frac{i}{4}\left(|u|^2+|v|^2\right) \\ \end{array} \right)
-\frac{i}{2z}\left( \begin{array}{cc} \bar{u}v & -\bar{u} \\
v+\bar{u}v^2 &  - \bar{u}v \\ \end{array} \right)\triangleq \mathcal{U}_1-\frac{i}{2z}\mathcal{U}_2,
 \\
\mathcal{V}=\left( \begin{array}{cc} -\frac{i}{4}\left(|u|^2-|v|^2\right) & -\frac{i}{2}\bar{v} \\
v_t-\frac{i}{2}|u|^2v-\frac{i}{2}u &  \frac{i}{4}\left(|u|^2-|v|^2\right) \\ \end{array} \right)
+\frac{i}{2z}\left( \begin{array}{cc} \bar{u}v & -\bar{u} \\
v+\bar{u}v^2 &  - \bar{u}v \\ \end{array} \right)\triangleq \mathcal{V}_1+\frac{i}{2z}\mathcal{U}_2.
\end{split}
\label{UV2}
\end{eqnarray}

\section{Direct part of the problem}

We assume that a solution $(u(x,t),v(x,t))$ exists and has sufficient smoothness and decay.
By performing the simultaneous spectral analysis of the Lax pair (\ref{LPk}) and (\ref{LPz}) respectively,
we will formulate two $2 \times 2$ matrix RH problems:
one problem is formulated in the complex $k$-plane and the other problem is formulated in the complex $z$-plane.
Each RH problem has explicit $(x,t)$-dependence and
is uniquely defined in terms of two sets of functions referred to as spectral functions.
The first set of functions is defined in terms of the initial data $(u_0(x),v_0(x))$,
whereas the remaining set is defined in terms of the Dirichlet boundary values $(g_0(t),h_0(t))$.
We will express $(u(x,t),v(x,t))$ in terms of the solutions of both RH problems.
The first expression is convenient for the reconstruction of the component $u(x,t)$
and the second one is convenient for the reconstruction of the component $v(x,t)$.

\subsection{Eigenfunctions}
For known $(u(x,t),v(x,t))$, the Lax pair (\ref{LPk}) can be considered as two linear equations for $\Phi(x,t,k)$.
Similarly, the Lax pair (\ref{LPz}) can be considered as two linear equations for $\Psi(x,t,z)$.
We want to construct respectively sectionally holomorphic solutions $\Phi(x,t,k)$ and $\Psi(x,t,z)$ in terms of $(u(x,t),v(x,t))$.

\subsubsection{Eigenfunctions $\{\Phi_j(x,t,k)\}_1^3$}
We define three different eigenfunctions $\{\Phi_j(x,t,k)\}_{1}^3$ as simultaneous solutions
of both parts of the Lax pair (\ref{LPk}) with normalization at $(x,t)=(0,\infty)$, at $(x,t)=(0,0)$, and at $(x,t)=(\infty,t)$, respectively.
These eigenfunctions are given by the following linear integral equations:
\begin{eqnarray}
\begin{split}
\Phi_{1}(x,t,k)=&I+\int_{0}^{x}e^{\frac{i}{4}[(k^{-1}-k)(x-\xi)]\hat{\sigma}_3}\left(U(\xi,t,k)\Phi_1(\xi,t,k)\right)d\xi
\\&-\int_{t}^{\infty}e^{\frac{i}{4}[(k^{-1}-k)x+(k^{-1}+k)(t-\tau)]\hat{\sigma}_3}\left(V(0,\tau,k)\Phi_1(0,\tau,k)\right)d\tau,
\\
\Phi_{2}(x,t,k)=&I+\int_{0}^{x}e^{\frac{i}{4}[(k^{-1}-k)(x-\xi)]\hat{\sigma}_3}\left(U(\xi,t,k)\Phi_2(\xi,t,k)\right)d\xi
\\&+\int_{0}^{t}e^{\frac{i}{4}[(k^{-1}-k)x+(k^{-1}+k)(t-\tau)]\hat{\sigma}_3}\left(V(0,\tau,k)\Phi_2(0,\tau,k)\right)d\tau,
\\
\Phi_{3}(x,t,k)=&I-\int_{x}^{\infty}e^{\frac{i}{4}[(k^{-1}-k)(x-\xi)]\hat{\sigma}_3}\left(U(\xi,t,k)\Phi_3(\xi,t,k)\right)d\xi.
\end{split}
\label{Phij}
\end{eqnarray}
For each $j=1,2,3$, we denote by $\Phi_j^L(x,t,k)$ and $\Phi_j^R(x,t,k)$ the first and second columns of $\Phi_j(x,t,k)$, respectively.
We introduce the domains in the complex $k$-plane as follows (see figure 1)
\begin{eqnarray}
\begin{split}
D_{1}=\left\{k\in\mathbb{C}, ~ |k|<1 \cap \text{Im}~ k>0\right\},
\\
D_{2}=\left\{k\in\mathbb{C}, ~ |k|>1 \cap \text{Im}~ k>0\right\},
\\
D_{3}=\left\{k\in\mathbb{C}, ~ |k|>1 \cap\text{Im}~ k<0\right\},
\\
D_{4}=\left\{k\in\mathbb{C}, ~ |k|<1 \cap \text{Im}~ k<0\right\},
\\
D_{-}=\left\{k\in\mathbb{C}, ~ \text{Im}~ k<0\right\},
\\
D_{+}=\left\{k\in\mathbb{C}, ~ \text{Im}~ k>0\right\},
\end{split}
\label{Domaink}
\end{eqnarray}
and we denote by $\overline{D}_{j}$, $j=1,2,3,4$, and $\overline{D}_{\pm}$ the closure of the above domains.
The eigenfunctions $\{\Phi_j(x,t,k)\}_{1}^3$ are bounded and analytic in the following domains:
\begin{eqnarray}
\begin{split}
\Phi_{1}^L(x,t,k): ~ k\in D_2;\quad \Phi_{1}^R(x,t,k): ~k\in D_3;
\\
\Phi_{2}^L(x,t,k): ~ k\in D_1;\quad \Phi_{2}^R(x,t,k): ~k\in D_4;
\\
\Phi_{3}^L(x,t,k): ~ k\in D_{-};\quad \Phi_{3}^R(x,t,k): ~k\in D_{+}.
\end{split}
\label{DE}
\end{eqnarray}
We can derive the following asymptotic behavior (see appendix A for details), for $j=1,2,3$,
\begin{eqnarray}
\Phi_j(x,t,k)=\left( \begin{array}{cc} \left(\Phi_j^0\right)^{11} & 0  \\
  0 & \left(\Phi_j^0\right)^{22} \\ \end{array} \right)
  +\frac{1}{k}\left( \begin{array}{cc} \left(\Phi_j^1\right)^{11} & \left(\Phi_j^1\right)^{12}  \\
  \left(\Phi_j^1\right)^{21} & \left(\Phi_j^1\right)^{22} \\ \end{array} \right)
  +O(\frac{1}{k^2}),
~~k\rightarrow \infty,
\label{aspPhi}
\end{eqnarray}
where
\begin{subequations}
\begin{eqnarray}
\left(\Phi_3^0\right)^{11}=\exp\left(\frac{i}{4}\int_x^\infty\left(|u(\xi,t)|^2+|v(\xi,t)|^2\right)d\xi\right),
\label{aspPhi1a}
\\
\left(\Phi_2^0\right)^{11}=\exp\left(-\frac{i}{4}\int_0^x\left(|u(\xi,t)|^2+|v(\xi,t)|^2\right)d\xi+\frac{i}{4}\int_0^t\left(|u(0,\tau)|^2-|v(0,\tau)|^2\right)d\tau\right),
\label{aspPhi1b}
\\
\left(\Phi_1^0\right)^{11}=\exp\left(-\frac{i}{4}\int_0^x\left(|u(\xi,t)|^2+|v(\xi,t)|^2\right)d\xi-\frac{i}{4}\int_t^\infty\left(|u(0,\tau)|^2-|v(0,\tau)|^2\right)d\tau\right),
\label{aspPhi1c}
\\
\left(\Phi_j^0\right)^{22}=\frac{1}{\left(\Phi_j^0\right)^{11}},
~~j=1,2,3,
\label{aspPhi1d}
\\
\left(\Phi_j^1\right)^{12}=\bar{u}\left(\Phi_j^0\right)^{22},
~~j=1,2,3,
\label{aspPhi1e}
\\
\left(\Phi_j^1\right)^{21}=\left(2iu_x+u|v|^2+v\right)\left(\Phi_j^0\right)^{11}=-\left(2iu_t+u|v|^2+v\right)\left(\Phi_j^0\right)^{11},
~~j=1,2,3.
\label{aspPhi1f}
\end{eqnarray}
\label{aspPhi1}
\end{subequations}

\begin{figure}
\begin{minipage}[t]{0.5\linewidth}
\centering
\includegraphics[width=2.5in]{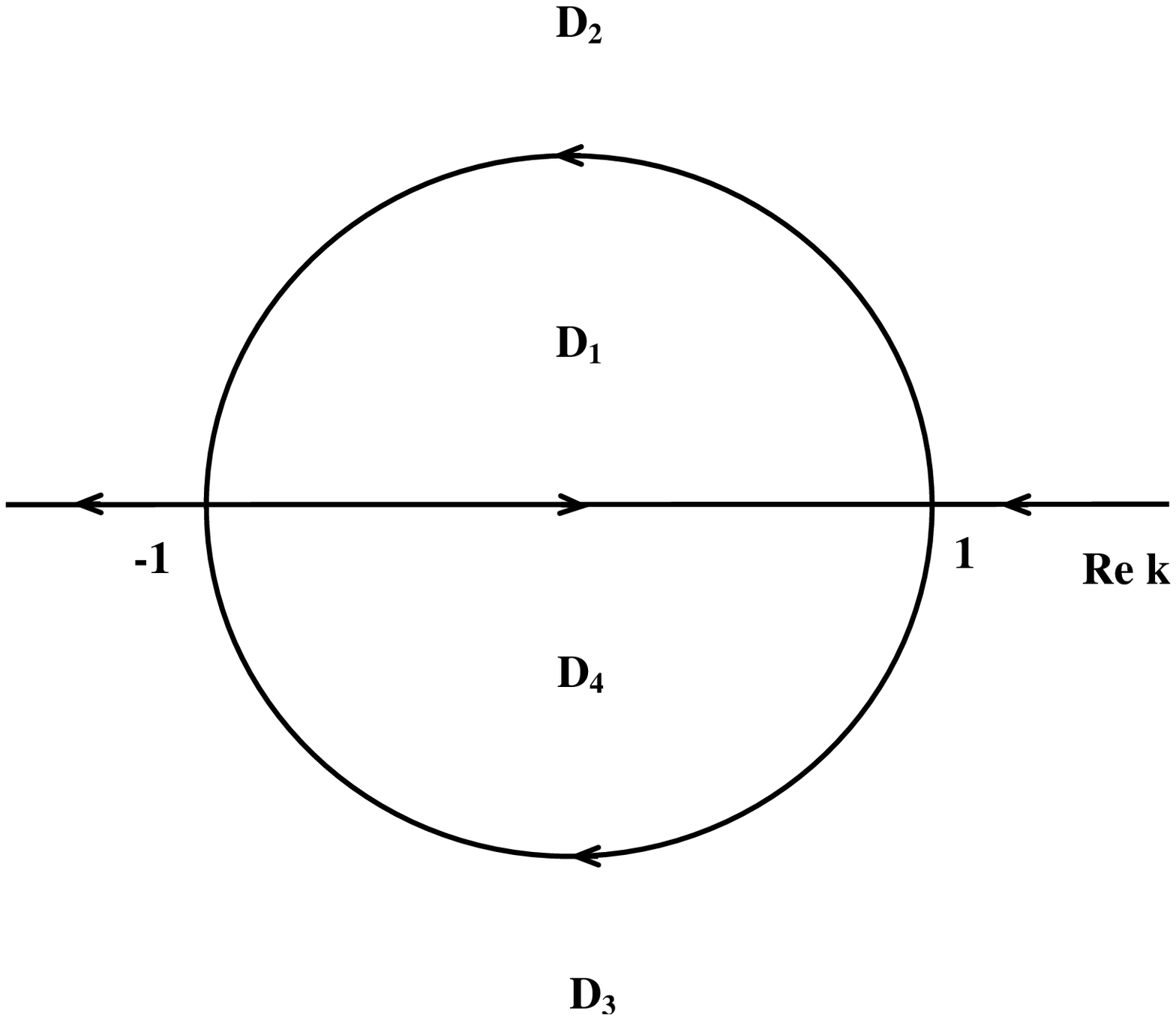}
\caption{\small{ The domains $\left\{D_{j}\right\}_1^4$ and contour $L$ in the complex $k$-plane.}}
\label{F1}
\end{minipage}
\hspace{2.0ex}
\begin{minipage}[t]{0.5\linewidth}
\centering
\includegraphics[width=2.5in]{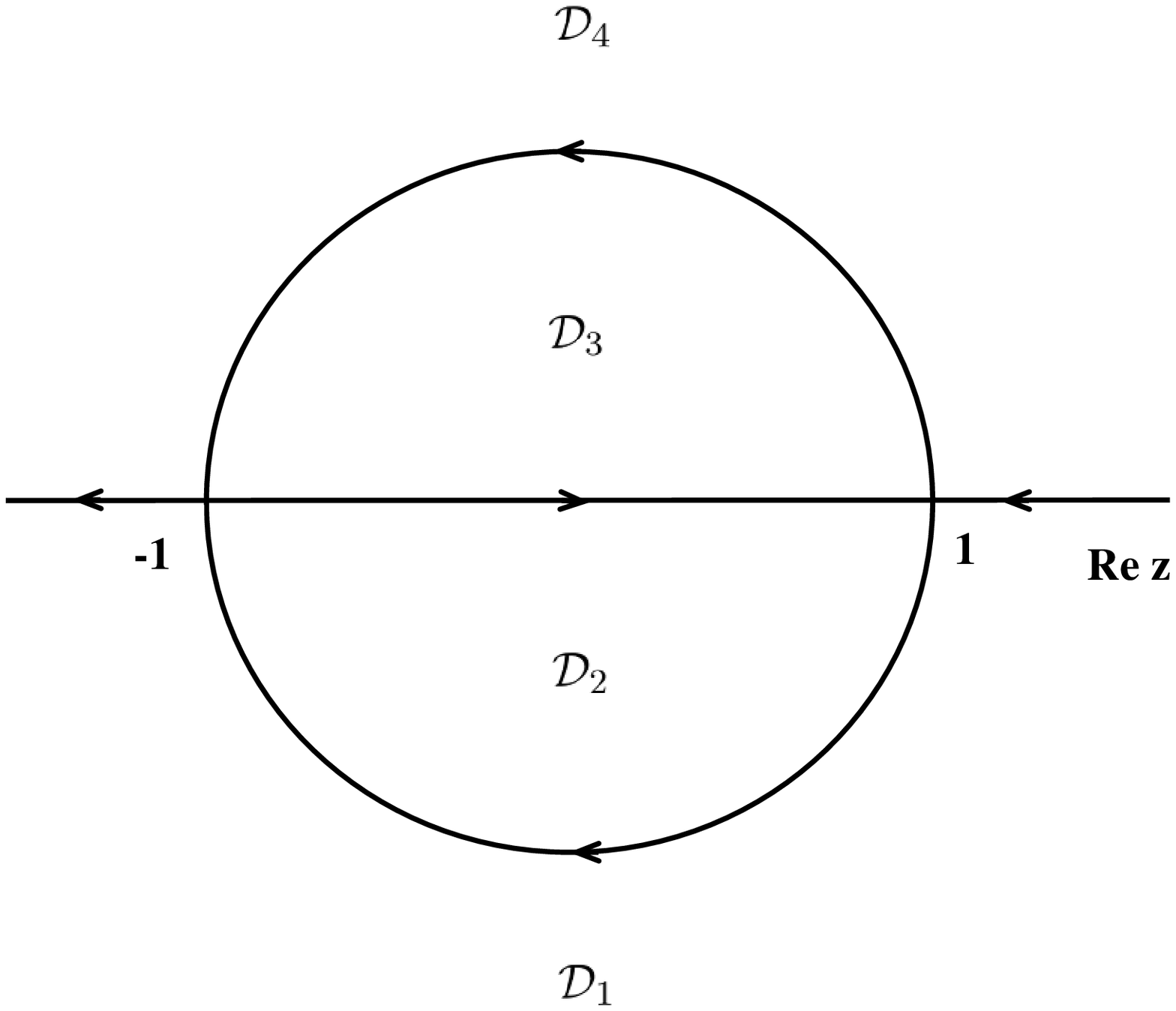}
\caption{\small{ The domains $\left\{\mathcal{D}_{j}\right\}_1^4$ and contour $\mathcal{L}$ in the complex $z$-plane.}}
\label{F2}
\end{minipage}
\end{figure}

\subsubsection{Eigenfunctions $\{\Psi_j(x,t,z)\}_1^3$}

We define three different eigenfunctions $\{\Psi_j(x,t,z)\}_{1}^3$ as simultaneous solutions
of both parts of the Lax pair (\ref{LPz}) with normalization at $(x,t)=(0,\infty)$, at $(x,t)=(0,0)$, and at $(x,t)=(\infty,t)$, respectively.
These eigenfunctions satisfy the following linear integral equations:
\begin{eqnarray}
\begin{split}
\Psi_{1}(x,t,z)=&I+\int_{0}^{x}e^{\frac{i}{4}[(z-z^{-1})(x-\xi)]\hat{\sigma}_3}\left(\mathcal{U}(\xi,t,z)\Psi_1(\xi,t,z)\right)d\xi
\\&-\int_{t}^{\infty}e^{\frac{i}{4}[(z-z^{-1})x+(z+z^{-1})(t-\tau)]\hat{\sigma}_3}\left(\mathcal{V}(0,\tau,z)\Psi_1(0,\tau,z)\right)d\tau,
\\
\Psi_{2}(x,t,z)=&I+\int_{0}^{x}e^{\frac{i}{4}[(z-z^{-1})(x-\xi)]\hat{\sigma}_3}\left(\mathcal{U}(\xi,t,z)\Psi_2(\xi,t,z)\right)d\xi
\\&+\int_{0}^{t}e^{\frac{i}{4}[(z-z^{-1})x+(z+z^{-1})(t-\tau)]\hat{\sigma}_3}\left(\mathcal{V}(0,\tau,z)\Psi_2(0,\tau,z)\right)d\tau,
\\
\Psi_{3}(x,t,z)=&I-\int_{x}^{\infty}e^{\frac{i}{4}[(z-z^{-1})(x-\xi)]\hat{\sigma}_3}\left(\mathcal{U}(\xi,t,z)\Psi_3(\xi,t,z)\right)d\xi.
\end{split}
\label{Psij}
\end{eqnarray}
As before, we denote by $\Psi_j^L(x,t,z)$ and $\Psi_j^R(x,t,z)$ the first and second columns of $\Psi_j(x,t,z)$, respectively.
We introduce the domains in the complex $z$-plane as follows (see figure 2)
\begin{eqnarray}
\begin{split}
\mathcal{D}_{1}=\left\{z\in\mathbb{C}, ~ |z|>1 \cap\text{Im}~ z<0\right\},
\\
\mathcal{D}_{2}=\left\{z\in\mathbb{C}, ~ |z|<1 \cap \text{Im}~ z<0\right\},
\\
\mathcal{D}_{3}=\left\{z\in\mathbb{C}, ~ |z|<1 \cap \text{Im}~ z>0\right\},
\\
\mathcal{D}_{4}=\left\{z\in\mathbb{C}, ~ |z|>1 \cap \text{Im}~ z>0\right\},
\\
\mathcal{D}_{+}=\left\{z\in\mathbb{C}, ~ \text{Im}~ z>0\right\},
\\
\mathcal{D}_{-}=\left\{z\in\mathbb{C}, ~ \text{Im}~ z<0\right\},
\end{split}
\label{Domainz}
\end{eqnarray}
and we denote by $\overline{\mathcal{D}}_{j}$, $j=1,2,3,4$, and $\overline{\mathcal{D}}_{\pm}$ the closure of these domains.
The eigenfunctions $\{\Psi_j(x,t,z)\}_{1}^3$ are bounded and analytic in the following domains:
\begin{eqnarray}
\begin{split}
\Psi_{1}^L(x,t,z): ~ z\in \mathcal{D}_2;\quad \Psi_{1}^R(x,t,z): ~z\in \mathcal{D}_3;
\\
\Psi_{2}^L(x,t,z): ~ z\in \mathcal{D}_1;\quad \Psi_{2}^R(x,t,z): ~z\in \mathcal{D}_4;
\\
\Psi_{3}^L(x,t,z): ~ z\in \mathcal{D}_{+};\quad \Psi_{3}^R(x,t,z): ~z\in \mathcal{D}_{-}.
\end{split}
\label{DEPsi}
\end{eqnarray}
With a similar derivation as shown in appendix A, we find the following asymptotic behavior, for $j=1,2,3$,
\begin{eqnarray}
\Psi_j(x,t,z)=\left( \begin{array}{cc} \left(\Psi_j^0\right)^{11} & 0  \\
  0 & \left(\Psi_j^0\right)^{22} \\ \end{array} \right)
  +\frac{1}{z}\left( \begin{array}{cc} \left(\Psi_j^1\right)^{11} & \left(\Psi_j^1\right)^{12}  \\
  \left(\Psi_j^1\right)^{21} & \left(\Psi_j^1\right)^{22} \\ \end{array} \right)
  +O(\frac{1}{z^2}),
~~z\rightarrow \infty,
\label{aspPsi}
\end{eqnarray}
where
\begin{eqnarray}
\begin{split}
\left(\Psi_3^0\right)^{11}=\exp\left(-\frac{i}{4}\int_x^\infty\left(|u(\xi,t)|^2+|v(\xi,t)|^2\right)d\xi\right),
\\
\left(\Psi_2^0\right)^{11}=\exp\left(\frac{i}{4}\int_0^x\left(|u(\xi,t)|^2+|v(\xi,t)|^2\right)d\xi-\frac{i}{4}\int_0^t\left(|u(0,\tau)|^2-|v(0,\tau)|^2\right)d\tau\right),
\\
\left(\Psi_1^0\right)^{11}=\exp\left(\frac{i}{4}\int_0^x\left(|u(\xi,t)|^2+|v(\xi,t)|^2\right)d\xi+\frac{i}{4}\int_t^\infty\left(|u(0,\tau)|^2-|v(0,\tau)|^2\right)d\tau\right),
\\
\left(\Psi_j^0\right)^{22}=\frac{1}{\left(\Psi_j^0\right)^{11}},
~~j=1,2,3,
\\
~~\left(\Psi_j^1\right)^{12}=\bar{v}\left(\Psi_j^0\right)^{22},
~~j=1,2,3,
\\
\left(\Psi_j^1\right)^{21}=\left(-2iv_x+|u|^2v+u\right)\left(\Psi_j^0\right)^{11}=-\left(2iv_t+|u|^2v+u\right)\left(\Psi_j^0\right)^{11},
~~j=1,2,3.
\end{split}
\label{aspPsi1}
\end{eqnarray}

\subsection{Spectral functions}

\subsubsection{Spectral functions in the complex $k$-plane}

Since the matrices $\{\Phi_j(x,t,k)\}_1^3$ satisfy the same linear equations (\ref{LPk}),
they are related:
\begin{subequations}
\begin{eqnarray}
\Phi_{3}(x,t,k)=\Phi_{2}(x,t,k)e^{\theta(x,t,k)\hat{\sigma_3}}s(k), \label{s}
\\
\Phi_{1}(x,t,k)=\Phi_{2}(x,t,k)e^{\theta(x,t,k)\hat{\sigma_3}}S(k),\label{S}
\end{eqnarray}
\label{sS}
\end{subequations}
where
\begin{eqnarray}
\theta(x,t,k)=\frac{i}{4}\left((k^{-1}-k)x+(k^{-1}+k)t\right).
\label{theta}
\end{eqnarray}
The matrices $U$ and $V$ in (\ref{LPk}) are traceless, this implies
$$\det\Phi_j(x,t,k)=1,~~j=1,2,3.$$
Thus
\begin{eqnarray}
\det s(k)=\det S(k)=1.
\label{det}
\end{eqnarray}

Evaluating equations (\ref{sS}) at $x=t=0$, we find
\begin{subequations}
\begin{eqnarray}
s(k)=\Phi_3(0,0,k),
\label{s1}
\\
S(k)=\Phi_1(0,0,k).
\label{S1}
\end{eqnarray}
\label{sS1}
\end{subequations}
We introduce the following notations:
\begin{subequations}
\begin{eqnarray}
s(k)=\left( \begin{array}{cc} \widetilde{a}(k) & b(k)  \\
  \widetilde{b}(k)  & a(k) \\ \end{array} \right),
\label{sSma}
\\
S(k)=\left( \begin{array}{cc} \widetilde{A}(k) & B(k)  \\
  \widetilde{B}(k)  & A(k) \\ \end{array} \right).
\label{sSmb}
\end{eqnarray}
\label{sSm}
\end{subequations}
From equations (\ref{aspPhi}), (\ref{aspPhi1}) and (\ref{sS1}), we find that:
\begin{eqnarray}
\begin{split}
&a(k)=\exp\left(-\frac{i}{4}\int_0^\infty\left(|u_0(\xi)|^2+|v_0(\xi)|^2\right)d\xi\right)+O(\frac{1}{k}), ~b(k)=O(\frac{1}{k}),~ k\rightarrow \infty, ~k\in \overline{D}_{+};
\\
&\widetilde{a}(k)=\exp\left(\frac{i}{4}\int_0^\infty\left(|u_0(\xi)|^2+|v_0(\xi)|^2\right)d\xi\right)+O(\frac{1}{k}), ~\widetilde{b}(k)=O(\frac{1}{k}), ~k\rightarrow \infty, ~k\in \overline{D}_{-};
\\
&A(k)=\exp\left(\frac{i}{4}\int_0^\infty\left(|g_0(\tau)|^2-|h_0(\tau)|^2\right)d\tau\right)+O(\frac{1}{k}), ~B(z)=O(\frac{1}{k}), ~k\rightarrow \infty, ~k\in \overline{D}_{1}\cup \overline{D}_{3};
\\
&\widetilde{A}(k)=\exp\left(-\frac{i}{4}\int_0^\infty\left(|g_0(\tau)|^2-|h_0(\tau)|^2\right)d\tau\right)+O(\frac{1}{k}), ~\widetilde{B}(z)=O(\frac{1}{k}), ~k\rightarrow \infty, ~k\in \overline{D}_{2}\cup \overline{D}_{4}.
\end{split}
\label{aspabAB}
\end{eqnarray}

Using symmetry properties of $Q$ and $P$,
we can deduce that, if $\varphi(x,t,\lambda)$ solves (\ref{LP1}),
then so does $\sigma_2 \overline{\varphi(x,t,\bar{\lambda})} \sigma_2$,
where $\sigma_2=\left( \begin{array}{cc} 0 & -1 \\
 1 &  0 \\ \end{array} \right)$.
 Using this fact and the transform (\ref{T1}),
 we find, if $\Phi_j(x,t,k)$, $j=1,2,3$, solves (\ref{LPk}),
then so does $G \overline{\Phi_j(x,t,\bar{k})} C_j$,
where
\begin{eqnarray}
\begin{split}
G=\left( \begin{array}{cc} \bar{u} & -1 \\
k+|u|^2 &  -u \\ \end{array} \right),
 \\
C_1=C_3=k^{-1}\left( \begin{array}{cc} 0 & 1 \\
-k &  0 \\ \end{array} \right),
\\
C_2=k^{-1}\left( \begin{array}{cc} -u_0(0)e^{-2\theta(x,t,k)} & 1 \\
-k-|u_0(0)|^2 & \bar{u}_0(0) e^{2\theta(x,t,k)} \\ \end{array} \right).
\end{split}
\label{GC}
\end{eqnarray}
Moreover, for each $j=1,2,3$, $G \overline{\Phi_j(x,t,\bar{k})} C_j$ satisfies the same normalized condition as that for $\Phi_j(x,t,k)$.
The uniqueness of normalized solutions implies that
\begin{eqnarray}
\Phi_j(x,t,k)=G \overline{\Phi_j(x,t,\bar{k})} C_j, j=1,2,3,
\label{Phisym}
\end{eqnarray}
where $G$ and $C_j$ are defined by (\ref{GC}).
Using equations (\ref{sS1}) and (\ref{Phisym}), we find the following symmetry relations
\begin{subequations}
\begin{eqnarray}
\widetilde{a}(k)=\overline{a(\bar{k})}-\bar{u}_0(0) \overline{b(\bar{k})},
\label{symsfa}
\\
\widetilde{b}(k)=u_0(0)\overline{a(\bar{k})}-\overline{b(\bar{k})}\left(k+|u_0(0)|^2\right),
\label{symsfb}
\\
\widetilde{A}(k)=\overline{A(\bar{k})}-\bar{u}_0(0) \overline{B(\bar{k})},
\label{symsfc}
\\
\widetilde{B}(k)=u_0(0)\overline{A(\bar{k})}-\overline{B(\bar{k})}\left(k+|u_0(0)|^2\right).
\label{symsfd}
\end{eqnarray}
\label{symsf}
\end{subequations}

\subsubsection{Spectral functions in the complex $z$-plane}

Since the matrices $\{\Psi_j(x,t,z)\}_1^3$ satisfy the same linear equations (\ref{LPz}),
they are related:
\begin{subequations}
\begin{eqnarray}
\Psi_{3}(x,t,z)=\Psi_{2}(x,t,z)e^{\vartheta(x,t,z)\hat{\sigma_3}}\mathfrak{s}(z), \label{sz}
\\
\Psi_{1}(x,t,z)=\Psi_{2}(x,t,z)e^{\vartheta(x,t,z)\hat{\sigma_3}}\mathcal{S}(z),\label{Sz}
\end{eqnarray}
\label{sSz}
\end{subequations}
where
\begin{eqnarray}
\vartheta(x,t,z)=\frac{i}{4}\left((z-z^{-1})x+(z+z^{-1})t\right).
\label{thetaz}
\end{eqnarray}
The matrices $\mathcal{U}$ and $\mathcal{V}$ in (\ref{LPz}) are traceless, this implies
$\det\Psi_j(x,t,z)=1,~j=1,2,3$,
and thus
\begin{eqnarray}
\det \mathfrak{s}(z)=\det \mathcal{S}(z)=1.
\label{detz}
\end{eqnarray}

Evaluating equations (\ref{sSz}) at $x=t=0$, we find
\begin{subequations}
\begin{eqnarray}
\mathfrak{s}(z)=\Psi_3(0,0,z),
\label{s1z}
\\
\mathcal{S}(z)=\Psi_1(0,0,z).
\label{S1z}
\end{eqnarray}
\label{sS1z}
\end{subequations}
We introduce the following notations:
\begin{subequations}
\begin{eqnarray}
\mathfrak{s}(z)=\left( \begin{array}{cc} \widetilde{\mathfrak{a}}(z) & \mathfrak{b}(z)  \\
  \widetilde{\mathfrak{b}}(z)  & \mathfrak{a}(z) \\ \end{array} \right),
\label{sSmaz}
\\
\mathcal{S}(z)=\left( \begin{array}{cc} \widetilde{\mathcal{A}}(z) & \mathcal{B}(z)  \\
  \widetilde{\mathcal{B}}(z)  & \mathcal{A}(z) \\ \end{array} \right).
\label{sSmbz}
\end{eqnarray}
\label{sSmz}
\end{subequations}
From equations (\ref{aspPsi}) and (\ref{sS1z}), we find
\begin{eqnarray}
\begin{split}
&\mathfrak{a}(z)=\exp\left(\frac{i}{4}\int_0^\infty\left(|u_0(\xi)|^2+|v_0(\xi)|^2\right)d\xi\right)+O(\frac{1}{z}), ~\mathfrak{b}(z)=O(\frac{1}{z}), ~z\rightarrow \infty, ~z\in \overline{\mathcal{D}}_{-};
\\
&\widetilde{\mathfrak{a}}(z)=\exp\left(-\frac{i}{4}\int_0^\infty\left(|u_0(\xi)|^2+|v_0(\xi)|^2\right)d\xi\right)+O(\frac{1}{z}),
~\widetilde{\mathfrak{b}}(z)=O(\frac{1}{z}), ~z\rightarrow \infty, ~z\in \overline{\mathcal{D}}_{+};
\\
&\mathcal{A}(z)=\exp\left(-\frac{i}{4}\int_0^\infty\left(|g_0(\tau)|^2-|h_0(\tau)|^2\right)d\tau\right)+O(\frac{1}{z}),
~\mathcal{B}(z)=O(\frac{1}{z}), ~z\rightarrow \infty, z\in \overline{D}_{1}\cup \overline{\mathcal{D}}_{3};
\\
&\widetilde{\mathcal{A}}(z)=\exp\left(\frac{i}{4}\int_0^\infty\left(|g_0(\tau)|^2-|h_0(\tau)|^2\right)d\tau\right)+O(\frac{1}{z}),
~\widetilde{\mathcal{B}}(z)=O(\frac{1}{z}), ~z\rightarrow \infty, ~z\in \overline{D}_{2}\cup \overline{\mathcal{D}}_{4}.
\end{split}
\label{aspabAB1z}
\end{eqnarray}

Similar with the case of the functions $\left\{\Phi_j(x,t,k)\right\}_1^3$, we find the following symmetry relations for $\left\{\Psi_j(x,t,z)\right\}_1^3$:
\begin{eqnarray}
\Psi_j(x,t,z)=\mathcal{G} \overline{\Psi_j(x,t,\bar{z})} \mathcal{C}_j, j=1,2,3,
\label{Psisymz}
\end{eqnarray}
where
\begin{eqnarray}
\begin{split}
\mathcal{G}=\left( \begin{array}{cc} z^{-1}\bar{v} & -z^{-1} \\
1+z^{-1}|v|^2 &  -z^{-1}v \\ \end{array} \right),
 \\
\mathcal{C}_1=\mathcal{C}_3=\left( \begin{array}{cc} 0 & 1 \\
-z &  0 \\ \end{array} \right),
\\
\mathcal{C}_2=\left( \begin{array}{cc} -v_0(0)e^{-2\vartheta(x,t,z)} & 1 \\
-z-|v_0(0)|^2 & \bar{v}_0(0) e^{2\vartheta(x,t,z)} \\ \end{array} \right).
\end{split}
\label{GCz}
\end{eqnarray}
Equations (\ref{sS1z}) and (\ref{Psisymz}) imply the following symmetry relations:
\begin{subequations}
\begin{eqnarray}
\widetilde{\mathfrak{a}}(z)=\overline{\mathfrak{a}(\bar{z})}-\bar{v}_0(0) \overline{\mathfrak{b}(\bar{z})},
\label{symsfaz}
\\
\widetilde{\mathfrak{b}}(z)=v_0(0)\overline{\mathfrak{a}(\bar{z})}-\overline{\mathfrak{b}(\bar{z})}\left(z+|v_0(0)|^2\right),
\label{symsfbz}
\\
\widetilde{\mathcal{A}}(z)=\overline{\mathcal{A}(\bar{z})}-\bar{v}_0(0) \overline{\mathcal{B}(\bar{z})},
\label{symsfcz}
\\
\widetilde{\mathcal{B}}(z)=v_0(0)\overline{\mathcal{A}(\bar{z})}-\overline{\mathcal{B}(\bar{z})}\left(z+|v_0(0)|^2\right).
\label{symsfdz}
\end{eqnarray}
\label{symsfz}
\end{subequations}

\subsection{Riemann-Hilbert problems}\label{RHsection}

\subsubsection{Riemann-Hilbert problem in the complex $k$-plane}

We define the matrices $M_{+}(x,t,k)$ and $M_{-}(x,t,k)$ as follows:
\begin{eqnarray}
\begin{split}
M_{+}(x,t,k)=\left\{\begin{array}{l}E(x,t)\left(\frac{\Phi_2^L(x,t,k)}{a(k)},\Phi_3^R(x,t,k)\right), \quad k\in \overline{D}_1,
\vspace{3mm}
\\
E(x,t)\left(\Phi_3^L(x,t,k),\frac{\Phi_1^R(x,t,k)}{\widetilde{d}(k)}\right), \quad k\in \overline{D}_3,
\vspace{3mm}
\end{array}\right.
 \\
M_{-}(x,t,k)=\left\{\begin{array}{l}E(x,t)\left(\frac{\Phi_1^L(x,t,k)}{d(k)},\Phi_3^R(x,t,k)\right), \quad k\in \overline{D}_2,
\vspace{3mm}
\\
E(x,t)\left(\Phi_3^L(x,t,k),\frac{\Phi_2^R(x,t,k)}{\widetilde{a}(k)}\right), \quad k\in \overline{D}_4,
\vspace{3mm}
\end{array}\right.
\end{split}
\label{M1}
\end{eqnarray}
where
\begin{eqnarray}
d(k)=a(k)\widetilde{A}(k)-b(k)\widetilde{B}(k), ~~\widetilde{d}(k)=\widetilde{a}(k)A(k)-\widetilde{b}(k)B(k),
\label{d}
\end{eqnarray}
and
\begin{eqnarray}
E(x,t)=\text{diag}\left(\exp\left(-\frac{i}{4}\int_x^\infty\left(|u(\xi,t)|^2+|v(\xi,t)|^2\right)d\xi\right),
\exp\left(\frac{i}{4}\int_x^\infty\left(|u(\xi,t)|^2+|v(\xi,t)|^2\right)d\xi\right)\right).
\label{E}
\end{eqnarray}

From (\ref{aspPhi}), (\ref{aspabAB}) and (\ref{M1}), we find
\begin{eqnarray}
\begin{split}
M(x,t,k)= I+M_1(x,t)\frac{1}{k}+O(\frac{1}{k^2}),  ~~ k\rightarrow \infty,
\end{split}
\label{Masp1}
\end{eqnarray}
where the $(1,2)$-entry and $(2,1)$-entry of the matrix $M_1(x,t)$ are given by
\begin{eqnarray}
\begin{split}
M_1^{12}(x,t)=\bar{u}\exp\left(-\frac{i}{2}\int_x^\infty\left(|u(\xi,t)|^2+|v(\xi,t)|^2\right)d\xi\right),
\\
M_1^{21}(x,t)=\left(2iu_x+u|v|^2+v\right)\exp\left(\frac{i}{2}\int_x^\infty\left(|u(\xi,t)|^2+|v(\xi,t)|^2\right)d\xi\right).
\end{split}
\label{M112}
\end{eqnarray}

By direct algebraic manipulations, equations (\ref{sS}) can be rewritten in the form of the jump condition of a RH problem:
\begin{eqnarray}
M_{-}(x,t,k)=M_{+}(x,t,k)J(x,t,k), \quad k\in L,
\label{RHP1}
\end{eqnarray}
where the jump matrix $J(x,t,k)$ and the oriented contour $L=L_1\cup L_2\cup L_3\cup L_4$ are defined as follows
\begin{eqnarray}
\begin{split}
J(x,t,k)&=\left\{\begin{array}{l}
J_1(x,t,k), \quad k\in \overline{D}_1\cap \overline{D}_2 \triangleq L_1,
\\
J_2(x,t,k), \quad k\in \overline{D}_2\cap \overline{D}_3 \triangleq L_2,
\\
J_3(x,t,k), \quad k\in \overline{D}_3\cap \overline{D}_4 \triangleq L_3,
\\
J_4(x,t,k), \quad k\in \overline{D}_1\cap \overline{D}_4 \triangleq L_4,
\end{array}\right.
 \end{split}
\label{Jm1}
\end{eqnarray}
\begin{eqnarray}
\begin{split}
J_1=e^{\theta(x,t,k)\hat{\sigma_3}}\left( \begin{array}{cc} 1 & 0 \\
  \Gamma(k) & 1 \\ \end{array} \right),
\\
J_3=e^{\theta(x,t,k)\hat{\sigma_3}}\left( \begin{array}{cc} 1 & -\widetilde{\Gamma}(k) \\
  0 & 1 \\ \end{array} \right),
\\
J_4=e^{\theta(x,t,k)\hat{\sigma_3}}\left( \begin{array}{cc} 1 & -\gamma(k) \\
  \widetilde{\gamma}(k) & 1-\gamma(k)\widetilde{\gamma}(k) \\ \end{array} \right),
\\
J_2=J_3J_4^{-1}J_1,
\end{split}
\label{JM1j}
\end{eqnarray}
with
\begin{eqnarray}
\begin{split}
\gamma(k)=\frac{b(k)}{\widetilde{a}(k)}, ~k\in \mathbb{R};
~~\widetilde{\gamma}(k)=\frac{\widetilde{b}(k)}{a(k)}, ~k\in \mathbb{R};
\\
\Gamma(k)=\frac{\widetilde{B}(k)}{a(k)d(k)}, ~k\in \overline{D}_2;
~~\widetilde{\Gamma}(k)=\frac{B(k)}{\widetilde{a}(k)\widetilde{d}(k)},~k\in \overline{D}_3.
\end{split}
\label{gG}
\end{eqnarray}
See figure 1 for the contour $L$ for this RH problem.

Assume that
\begin{itemize}
\item $a(k)$ has at most $n$ simple zeros $\{k_j\}_{1}^{n}$, $n=n_1+n_2$, where $k_j\in D_{1}$, $j=1,\cdots,n_1$; $k_j\in D_{2}$, $j=n_1+1,\cdots,n_1+n_2$.
 $\widetilde{a}(k)$ has at most $\widetilde{n}$ simple zeros $\{\widetilde{k}_j\}_{1}^{\widetilde{n}}$, $\widetilde{n}=\widetilde{n}_1+\widetilde{n}_2$, where $\widetilde{k}_j\in D_{4}$, $j=1,\cdots,\widetilde{n}_1$; $\widetilde{k}_j\in D_{3}$, $j=\widetilde{n}_1+1,\cdots,\widetilde{n}_1+\widetilde{n}_2$.
\item  $d(k)$ has at most $K$ simple zeros $\{\kappa_j\}_{1}^{K}$, where $\kappa_j\in D_{2}$, $j=1,\cdots,K$.
  $\widetilde{d}(k)$ has at most $\widetilde{K}$ simple zeros $\{\widetilde{\kappa}_j\}_{1}^{\widetilde{K}}$, where $\widetilde{\kappa}_j\in D_{3}$, $j=1,\cdots,\widetilde{K}$.
\item  None of the zeros of $a(k)$ for $k\in D_{2}$ coincides with any of the zeros of $d(k)$.
   None of the zeros of $\widetilde{a}(k)$ for $k\in D_{3}$ coincides with any of the zeros of $\widetilde{d}(k)$.
\end{itemize}
From (\ref{sS}) and (\ref{M1}), we find the following residues conditions:
\begin{subequations}
\begin{eqnarray}
{\text{Res}}_{k=k_j}M^{L}(x,t,k)=\frac{e^{-2\theta(x,t,k_j)}}{\dot{a}(k_j)b(k_j)}M^{R}(x,t,k_j),~~j=1,\cdots,n_1,
\label{rr1a}
\\
{\text{Res}}_{k=\widetilde{k}_j}M^{R}(x,t,k)=\frac{e^{2\theta(x,t,\widetilde{k}_j)}}{\dot{\widetilde{a}}(\widetilde{k}_j)\widetilde{b}(\widetilde{k}_j)}
M^{L}(x,t,\widetilde{k}_j),~~j=1,\cdots,\widetilde{n}_1,
\label{rr1b}
\\
{\text{Res}}_{k=\kappa_j}M^{L}(x,t,k)=\frac{\widetilde{B}(\kappa_j)e^{-2\theta(x,t,\kappa_j)}}{a(\kappa_j)\dot{d}(\kappa_j)}M^{R}(x,t,\kappa_j),~~j=1,\cdots,K,
\label{rr1c}
\\
{\text{Res}}_{k=\widetilde{\kappa}_j}M^{R}(x,t,k)=\frac{B(\widetilde{\kappa}_j)e^{2\theta(x,t,\widetilde{\kappa}_j)}}{\widetilde{a}(\widetilde{\kappa}_j)
\dot{\widetilde{d}}(\widetilde{\kappa}_j)}
M^{L}(x,t,\widetilde{\kappa}_j),~~j=1,\cdots,\widetilde{K},
\label{rr1d}
\end{eqnarray}
\label{rr1}
\end{subequations}
where $\theta(x,t,k_j)=\frac{i}{4}\left((k_j^{-1}-k_j)x+(k_j^{-1}+k_j)t\right)$ and $\dot{a}(k)=\frac{da}{dk}$.

Finally, we note that the asymptotic behavior (\ref{Masp1}) with (\ref{M112}) implies the following expressions
for $u(x,t)$ and $v(x,t)$ in terms of $M(x,t,k)$:
\begin{eqnarray}
\begin{split}
\bar{u}\exp\left(-\frac{i}{2}\int_x^\infty\left(|u(\xi,t)|^2+|v(\xi,t)|^2\right)d\xi\right)
=\lim_{k\rightarrow \infty}\left(k M^{12}(x,t,k)\right),
\\
\left(2iu_x+u|v|^2+v\right)\exp\left(\frac{i}{2}\int_x^\infty\left(|u(\xi,t)|^2+|v(\xi,t)|^2\right)d\xi\right)
=\lim_{k\rightarrow \infty}\left(kM^{21}(x,t,k)\right),
\end{split}
\label{uvs1}
\end{eqnarray}
where the subscripts $12$ and $21$ denote the $(1,2)$-entry and $(2,1)$-entry of a matrix.

\subsubsection{Riemann-Hilbert problem in the complex $z$-plane}

We define the matrices $\mathcal{M}_{+}(x,t,z)$ and $\mathcal{M}_{-}(x,t,z)$ as follows:
\begin{eqnarray}
\begin{split}
\mathcal{M}_{+}(x,t,z)=\left\{\begin{array}{l}\mathcal{E}(x,t)\left(\frac{\Psi_2^L(x,t,z)}{\mathfrak{a}(z)},\Psi_3^R(x,t,z)\right), \quad z\in \overline{\mathcal{D}}_1,
\vspace{3mm}
\\
\mathcal{E}(x,t)\left(\Psi_3^L(x,t,z),\frac{\Psi_1^R(x,t,z)}{\widetilde{\alpha}(z)}\right), \quad z\in \overline{\mathcal{D}}_3,
\vspace{3mm}
\end{array}\right.
 \\
\mathcal{M}_{-}(x,t,z)=\left\{\begin{array}{l}\mathcal{E}(x,t)\left(\frac{\Psi_1^L(x,t,z)}{\alpha(z)},\Psi_3^R(x,t,z)\right), \quad z\in \overline{\mathcal{D}}_2,
\vspace{3mm}
\\
\mathcal{E}(x,t)\left(\Psi_3^L(x,t,z),\frac{\Psi_2^R(x,t,z)}{\widetilde{\mathfrak{a}}(z)}\right), \quad z\in \overline{\mathcal{D}}_4,
\vspace{3mm}
\end{array}\right.
\end{split}
\label{M2}
\end{eqnarray}
where
\begin{eqnarray}
\alpha(z)=\mathfrak{a}(z)\widetilde{\mathcal{A}}(z)-\mathfrak{b}(z)\widetilde{\mathcal{B}}(z),
~~\widetilde{\alpha}(z)=\widetilde{\mathfrak{a}}(z)\mathcal{A}(z)-\widetilde{\mathfrak{b}}(z)\mathcal{B}(z),
\label{alpha}
\end{eqnarray}
and
\begin{eqnarray}
\mathcal{E}(x,t)=\text{diag}\left(\exp\left(\frac{i}{4}\int_x^\infty\left(|u(\xi,t)|^2+|v(\xi,t)|^2\right)d\xi\right),
\exp\left(-\frac{i}{4}\int_x^\infty\left(|u(\xi,t)|^2+|v(\xi,t)|^2\right)d\xi\right)\right).
\label{E2}
\end{eqnarray}

From (\ref{aspPsi}), (\ref{aspabAB1z}) and (\ref{M2}), we find
\begin{eqnarray}
\begin{split}
\mathcal{M}(x,t,z)= I+\mathcal{M}_1(x,t)\frac{1}{z}+O(\frac{1}{z^2}),  ~~ z\rightarrow \infty,
\end{split}
\label{Masp2}
\end{eqnarray}
where the $(1,2)$-entry and $(2,1)$-entry of the matrix $\mathcal{M}_1(x,t)$ are given by
\begin{eqnarray}
\begin{split}
\mathcal{M}_1^{12}(x,t)=\bar{v}\exp\left(\frac{i}{2}\int_x^\infty\left(|u(\xi,t)|^2+|v(\xi,t)|^2\right)d\xi\right),
\\
\mathcal{M}_1^{21}(x,t)=\left(-2iv_x+|u|^2v+u\right)\exp\left(-\frac{i}{2}\int_x^\infty\left(|u(\xi,t)|^2+|v(\xi,t)|^2\right)d\xi\right).
\end{split}
\label{M212}
\end{eqnarray}

By direct algebraic manipulations, equations (\ref{sSz}) can be rewritten in the form of the jump condition of a RH problem:
\begin{eqnarray}
\mathcal{M}_{-}(x,t,z)=\mathcal{M}_{+}(x,t,z)\mathcal{J}(x,t,z), \quad z\in \mathcal{L},
\label{RHP2}
\end{eqnarray}
where the jump matrix $\mathcal{J}(x,t,z)$ and the oriented contour $\mathcal{L}=\mathcal{L}_1\cup \mathcal{L}_2\cup \mathcal{L}_3\cup \mathcal{L}_4$
are defined as follows
\begin{eqnarray}
\begin{split}
\mathcal{J}(x,t,z)&=\left\{\begin{array}{l}
\mathcal{J}_1(x,t,z), \quad z\in \overline{\mathcal{D}}_1\cap \overline{\mathcal{D}}_2 \triangleq \mathcal{L}_1,
\\
\mathcal{J}_2(x,t,z), \quad z\in \overline{\mathcal{D}}_2\cap \overline{\mathcal{D}}_3 \triangleq \mathcal{L}_2,
\\
\mathcal{J}_3(x,t,z), \quad z\in \overline{\mathcal{D}}_3\cap \overline{\mathcal{D}}_4 \triangleq \mathcal{L}_3,
\\
\mathcal{J}_4(x,t,z), \quad z\in \overline{\mathcal{D}}_1\cap \overline{\mathcal{D}}_4 \triangleq \mathcal{L}_4,
\end{array}\right.
 \end{split}
\label{Jm2}
\end{eqnarray}
\begin{eqnarray}
\begin{split}
\mathcal{J}_1=e^{\vartheta(x,t,z)\hat{\sigma_3}}\left( \begin{array}{cc} 1 & 0 \\
  R(z) & 1 \\ \end{array} \right),
\\
\mathcal{J}_3=e^{\vartheta(x,t,z)\hat{\sigma_3}}\left( \begin{array}{cc} 1 & -\widetilde{R}(z) \\
  0 & 1 \\ \end{array} \right),
\\
\mathcal{J}_4=e^{\vartheta(x,t,z)\hat{\sigma_3}}\left( \begin{array}{cc} 1 & -r(z) \\
  \widetilde{r}(z) & 1-r(z)\widetilde{r}(z) \\ \end{array} \right),
\\
\mathcal{J}_2=\mathcal{J}_3\mathcal{J}_4^{-1}\mathcal{J}_1,
\end{split}
\label{JM2j}
\end{eqnarray}
with
\begin{eqnarray}
\begin{split}
r(z)=\frac{\mathfrak{b}(z)}{\widetilde{\mathfrak{a}}(z)}, ~z\in \mathbb{R};
~~\widetilde{r}(z)=\frac{\widetilde{\mathfrak{b}}(z)}{\mathfrak{a}(z)}, ~z\in \mathbb{R};
\\
R(z)=\frac{\widetilde{\mathcal{B}}(z)}{\mathfrak{a}(z)\alpha(z)}, ~z\in \overline{\mathcal{D}}_2;
~~\widetilde{R}(z)=\frac{\mathcal{B}(z)}{\widetilde{\mathfrak{a}}(z)\widetilde{\alpha}(z)},~z\in \overline{\mathcal{D}}_3.
\end{split}
\label{rR}
\end{eqnarray}
See figure 2 for the contour $\mathcal{L}$ for this RH problem.

Assume that
\begin{itemize}
\item $\mathfrak{a}(z)$ has at most $m$ simple zeros $\{z_j\}_{1}^{m}$, $m=m_1+m_2$, where $z_j\in \mathcal{D}_{1}$, $j=1,\cdots,m_1$; $z_j\in \mathcal{D}_{2}$, $j=m_1+1,\cdots,m_1+m_2$.
 $\widetilde{\mathfrak{a}}(z)$ has at most $\widetilde{m}$ simple zeros $\{\widetilde{z}_j\}_{1}^{\widetilde{m}}$, $\widetilde{m}=\widetilde{m}_1+\widetilde{m}_2$, where $\widetilde{z}_j\in \mathcal{D}_{4}$, $j=1,\cdots,\widetilde{m}_1$; $\widetilde{z}_j\in \mathcal{D}_{3}$, $j=\widetilde{m}_1+1,\cdots,\widetilde{m}_1+\widetilde{m}_2$.
\item  $\alpha(z)$ has at most $\Lambda$ simple zeros $\{\nu_j\}_{1}^{\Lambda}$, where $\nu_j\in \mathcal{D}_{2}$, $j=1,\cdots,\Lambda$.
  $\widetilde{\alpha}(z)$ has at most $\widetilde{\Lambda}$ simple zeros $\{\widetilde{\nu}_j\}_{1}^{\widetilde{\Lambda}}$, where $\widetilde{\nu}_j\in \mathcal{D}_{3}$, $j=1,\cdots,\widetilde{\Lambda}$.
\item  None of the zeros of $\mathfrak{a}(z)$ for $z\in \mathcal{D}_{2}$ coincides with any of the zeros of $\alpha(z)$.
   None of the zeros of $\widetilde{\mathfrak{a}}(z)$ for $z\in \mathcal{D}_{3}$ coincides with any of the zeros of $\widetilde{\alpha}(z)$.
\end{itemize}
From (\ref{sSz}) and (\ref{M2}), we find the following residues conditions:
\begin{subequations}
\begin{eqnarray}
{\text{Res}}_{z=z_j}\mathcal{M}^{L}(x,t,z)=\frac{e^{-2\vartheta(x,t,z_j)}}{\dot{\mathfrak{a}}(z_j)\mathfrak{b}(z_j)}\mathcal{M}^{R}(x,t,z_j),~~j=1,\cdots,m_1,
\label{rr2a}
\\
{\text{Res}}_{z=\widetilde{z}_j}\mathcal{M}^{R}(x,t,z)=\frac{e^{2\vartheta(x,t,\widetilde{z}_j)}}{\dot{\widetilde{\mathfrak{a}}}(\widetilde{z}_j)
\widetilde{\mathfrak{b}}(\widetilde{z}_j)}
\mathcal{M}^{L}(x,t,\widetilde{z}_j),~~j=1,\cdots,\widetilde{m}_1,
\label{rr2b}
\\
{\text{Res}}_{z=\nu_j}\mathcal{M}^{L}(x,t,z)=\frac{\widetilde{\mathcal{B}}(\nu_j)e^{-2\vartheta(x,t,\nu_j)}}{\mathfrak{a}(\nu_j)\dot{\alpha}(\nu_j)}
\mathcal{M}^{R}(x,t,\nu_j),~~j=1,\cdots,\Lambda,
\label{rr2c}
\\
{\text{Res}}_{z=\widetilde{\nu}_j}\mathcal{M}^{R}(x,t,z)=\frac{\mathcal{B}(\widetilde{\nu}_j)e^{2\vartheta(x,t,\widetilde{\nu}_j)}}
{\widetilde{\mathfrak{a}}(\widetilde{\nu}_j)\dot{\widetilde{\alpha}}(\widetilde{\nu}_j)}
\mathcal{M}^{L}(x,t,\widetilde{\nu}_j),~~j=1,\cdots,\widetilde{\Lambda},
\label{rr2d}
\end{eqnarray}
\label{rr2}
\end{subequations}
where $\vartheta(x,t,z_j)=\frac{i}{4}\left((z_j-z_j^{-1})x+(z_j+z_j^{-1})t\right)$ and $\dot{\mathfrak{a}}(z)=\frac{d\mathfrak{a}}{dz}$.

The asymptotic behavior (\ref{Masp2}), (\ref{M212}) implies the following expressions for $u(x,t)$ and $v(x,t)$:
\begin{eqnarray}
\begin{split}
\bar{v}\exp\left(\frac{i}{2}\int_x^\infty\left(|u(\xi,t)|^2+|v(\xi,t)|^2\right)d\xi\right)=\lim_{z\rightarrow \infty}\left(z\mathcal{M}^{12}(x,t,z)\right),
\\
\left(-2iv_x+|u|^2v+u\right)\exp\left(-\frac{i}{2}\int_x^\infty\left(|u(\xi,t)|^2+|v(\xi,t)|^2\right)d\xi\right)=\lim_{z\rightarrow \infty}\left(z\mathcal{M}^{21}(x,t,z)\right).
\end{split}
\label{uvs2}
\end{eqnarray}
where the subscripts $12$ and $21$ denote the $(1,2)$-entry and $(2,1)$-entry of a matrix.

\subsection{Global relations}

Using (\ref{S}) in (\ref{s}) to eliminate $\Phi_2(x,t,k)$, we find
\begin{eqnarray}
S^{-1}(k)s(k)=e^{-\theta(x,t,k)\hat{\sigma_3}}\left(\Phi_1^{-1}(x,t,k)\Phi_3(x,t,k)\right).
\label{gr1}
\end{eqnarray}
Evaluating the $(1,2)$-entry of (\ref{gr1})  at $x=0$ and $t\rightarrow \infty$,
we obtain the following global relation in the complex $k$-plane:
\begin{eqnarray}
A(k)b(k)-a(k)B(k)=0, ~~k\in \overline{D}_1.
\label{grk}
\end{eqnarray}

On the other hand, using (\ref{Sz}) in (\ref{sz}) to eliminate $\Psi_2(x,t,z)$, we find
\begin{eqnarray}
\mathcal{S}^{-1}(z)\mathfrak{s}(z)=e^{-\vartheta(x,t,z)\hat{\sigma_3}}\left(\Psi_1^{-1}(x,t,z)\Psi_3(x,t,z)\right).
\label{gr1z}
\end{eqnarray}
Evaluating the $(1,2)$-entry of (\ref{gr1z}) at $x=0$ and $t\rightarrow \infty$, we find the following global relation in the complex $z$-plane:
\begin{eqnarray}
\mathcal{A}(z)\mathfrak{b}(z)-\mathfrak{a}(z)\mathcal{B}(z)=0, ~~z\in \overline{\mathcal{D}}_1.
\label{grz}
\end{eqnarray}

\section{Inverse part of the problem}

Motivated by the above analysis, we define the spectral functions $a(k)$ and $b(k)$ ($\mathfrak{a}(z)$ and $\mathfrak{b}(z)$) in terms of the
initial data $(u_0(x),v_0(x))$, and define the spectral functions $A(k)$ and $B(k)$ ($\mathcal{A}(z)$ and $\mathcal{B}(z)$)
in terms of the boundary values $(g_0(t),h_0(t))$.
We assume that the boundary values are such that the spectral functions satisfy
the global relation. We also define $(u(x,t),v(x,t))$ in terms of the solutions of the RH problems
formulated in section \ref{RHsection}.
We then prove that the reconstruction formulae (\ref{uvs1}) and (\ref{uvs2}) for $(u(x,t),v(x,t))$ satisfy the MT system (\ref{MT}),
and furthermore satisfy the initial-boundary condition (\ref{IBV}).

\subsection{Inverse part for the problem in the complex $k$-plane}

The spectral analysis, performed for the Lax pair (\ref{LPk}) in the above section,
motivates the following definitions and properties for the spectral functions in the complex $k$-plane.

Let
\begin{eqnarray}
\begin{split}
U_0(x,k)=&\left( \begin{array}{cc} -\frac{i}{4}\left(|u_0(x)|^2+|v_0(x)|^2\right) & \frac{i}{2}\bar{u}_0(x)\\
u'_0(x)-\frac{i}{2}u_0(x)|v_0(x)|^2-\frac{i}{2}v_0(x) &  \frac{i}{4}\left(|u_0(x)|^2+|v_0(x)|^2\right) \\ \end{array} \right)
\\
&+\frac{i}{2k}\left( \begin{array}{cc} u_0(x) \bar{v}_0(x) & -\bar{v}_0(x) \\
u_0(x)+u^2_0(x)\bar{v}_0(x) &  -u_0(x) \bar{v}_0(x) \\ \end{array} \right)
\\
\triangleq
& \left( \begin{array}{cc} U_0^{11} & U_0^{12} \\
U_0^{21} &  -U_0^{11} \\ \end{array} \right),
\end{split}
\label{U0}
\end{eqnarray}
where $u'_0(x)=\frac{du_0(x)}{dx}$.
\begin{definition}
Given $(u_0(x),v_0(x))\in \mathrm{S}(\mathbb{R}^+)$ (here $\mathrm{S}(\mathbb{R}^+)$ denotes the functions of Schwartz class on $\mathbb{R}^+$),
we define the spectral functions $a(k)$ and $b(k)$ as follows:
\begin{eqnarray}
a(k)=\phi_2(0,k),~~b(k)=\phi_1(0,k), ~~{\rm Im} ~k\geq 0,
\label{dab}
\end{eqnarray}
where the vector function $\phi(x, k) = \left(\phi_1(x,k), \phi_2(x,k)\right)^T$ is the unique solution of
\begin{eqnarray}
\begin{split}
&\phi_{1,x}-\frac{i}{2}\left(k^{-1}-k\right)\phi_1=U_0^{11} \phi_1+U_0^{12} \phi_2,
\\
&\phi_{2,x}=U_0^{21} \phi_1-U_0^{11} \phi_2, ~~{\rm Im} ~k\geq 0, ~~0<x<\infty,
\\
&\lim_{x\rightarrow \infty}\phi=(0,1)^T.
\end{split}
\label{dabphi}
\end{eqnarray}
Moreover, we define $\widetilde{a}(k)$ and $\widetilde{b}(k)$ in terms of $a(k)$ and $b(k)$ via the symmetries (\ref{symsfa}) and (\ref{symsfb}).
\end{definition}

\begin{proposition}
The spectral functions $a(k)$ and $b(k)$ have the following properties:
\\(1) a(k) and b(k) are analytic for ${\rm Im} ~k>0$ and continuous and bounded for ${\rm Im} ~k\geq 0$.
\\(2) $a(k) = \exp\left(-\frac{i}{4}\int_0^\infty\left(|u_0(x)|^2+|v_0(x)|^2\right)dx\right) + O(\frac{1}{k})$, $b(k) =  O(\frac{1}{k})$, $k\rightarrow \infty$, ${\rm Im} ~k\geq 0.$
\\(3) $a(k)\widetilde{a}(k)-b(k)\widetilde{b}(k)=1$, $k \in \mathbb{R}.$
\\(4) $(u_0(x),v_0(x))$ can be reconstructed in terms of $a(k)$ and $b(k)$ by
\begin{eqnarray}
\begin{split}
\bar{u}_0(x)\exp\left(-\frac{i}{2}\int_x^\infty\left(|u_0(\xi)|^2+|v_0(\xi)|^2\right)d\xi\right)=\lim_{k\rightarrow \infty}(kM^{(x)}(x,k))^{12},
\\
(2iu'_0(x)+u_0(x)|v_0(x)|^2+v_0(x))\exp\left(\frac{i}{2}\int_x^\infty\left(|u_0(\xi)|^2+|v_0(\xi)|^2\right)d\xi\right)=\lim_{k\rightarrow \infty}(kM^{(x)}(x,k))^{21},
\end{split}
\label{abis}
\end{eqnarray}
where $M^{(x)}(x, k)$ is the unique solution of the following RH problem:
\\$\bullet$
\begin{eqnarray}
M^{(x)}(x, k)=\left\{ \begin{array}{cc} M_{-}^{(x)}(x, k), ~~{\rm Im} k\leq 0,   \\
  M_{+}^{(x)}(x, k), ~~{\rm Im} k\geq 0,  \\ \end{array} \right.
\label{abm}
\end{eqnarray}
is a sectionally meromorphic function.
\\$\bullet$
\begin{eqnarray}
M_{-}^{(x)}(x,k)=M_{+}^{(x)}(x,k)J^{(x)}(x,k), \quad k\in \mathbb{R},
\label{abJc}
\end{eqnarray}
where
\begin{eqnarray}
J^{(x)}(x,k)=e^{\frac{i}{4}(k^{-1}-k)x\hat{\sigma_3}}\left( \begin{array}{cc} 1 & -\gamma(k) \\
  \widetilde{\gamma}(k) & 1-\gamma(k)\widetilde{\gamma}(k) \\ \end{array} \right),
~\gamma(k)=\frac{b(k)}{\widetilde{a}(k)}, ~\widetilde{\gamma}(k)=\frac{\widetilde{b}(k)}{a(k)}.
\label{abJr}
\end{eqnarray}
\\$\bullet$
\begin{eqnarray}
M^{(x)}(x,k)=I+O(\frac{1}{k}), \quad k\rightarrow \infty.
\label{abMA}
\end{eqnarray}
\\$\bullet$
We assume that $a(k)$ can have at most $n$ simple zeros $\{k_j\}_{1}^{n}$, $n=n_1+n_2$, where $k_j\in D_{1}$, $j=1,\cdots,n_1$; $k_j\in D_{2}$, $j=n_1+1,\cdots,n_1+n_2$.
 $\widetilde{a}(k)$ can have at most $\widetilde{n}$ simple zeros $\{\widetilde{k}_j\}_{1}^{\widetilde{n}}$, $\widetilde{n}=\widetilde{n}_1+\widetilde{n}_2$, where $\widetilde{k}_j\in D_{4}$, $j=1,\cdots,\widetilde{n}_1$; $\widetilde{k}_j\in D_{3}$, $j=\widetilde{n}_1+1,\cdots,\widetilde{n}_1+\widetilde{n}_2$.
The first column of $M_{+}^{(x)}(x,k)$ can have simple poles at $k=k_j$, $j=1,\cdots,n$, and the second column of $M_{-}^{(x)}(x,k)$
can have simple poles at $k=\widetilde{k}_j$, $j=1,\cdots,\widetilde{n}$. The
associated residues are given by
\begin{subequations}
\begin{eqnarray}
{{\rm Res}}_{k=k_j}[M^{(x)}(x, k)]^L=\frac{e^{-\frac{i}{2}(k_j^{-1}-k_j)x}}{\dot{a}(k_j)b(k_j)}[M^{(x)}(x, k_j)]^R, ~~j=1,\cdots,n,
\label{abrr1a}
\\
{{\rm Res}}_{k=\widetilde{k}_j}[M^{(x)}(x, k)]^R=\frac{e^{\frac{i}{2}(\widetilde{k}_j^{-1}-\widetilde{k}_j)x}}{\dot{\widetilde{a}}(\widetilde{k}_j)\widetilde{b}(\widetilde{k}_j)}
[M^{(x)}(x,\widetilde{k}_j)]^{L},~~j=1,\cdots,\widetilde{n}.
\label{abrr1b}
\end{eqnarray}
\label{abrr1}
\end{subequations}
\end{proposition}
{\bf Proof} \quad Properties (1), (2) and (3) follow from definition 1.
The proof for property (4) is given in appendix B.  \QEDB

Let
\begin{eqnarray}
\begin{split}
V_0(t,k)=&\left( \begin{array}{cc} \frac{i}{4}\left(|g_0(t)|^2-|h_0(t)|^2\right) & -\frac{i}{2}\bar{g}_0(t)\\
g'_0(t)-\frac{i}{2}g_0(t)|h_0(t)|^2-\frac{i}{2}h_0(t) &  -\frac{i}{4}\left(|g_0(t)|^2-|h_0(t)|^2\right) \\ \end{array} \right)
\\
&+\frac{i}{2k}\left( \begin{array}{cc} g_0(t) \bar{h}_0(t) & -\bar{h}_0(t) \\
g_0(t)+g^2_0(t)\bar{h}_0(t) &  -g_0(t) \bar{h}_0(t) \\ \end{array} \right)
\\
\triangleq
& \left( \begin{array}{cc} V_0^{11} & V_0^{12} \\
V_0^{21} &  -V_0^{11} \\ \end{array} \right),
\end{split}
\label{V0}
\end{eqnarray}
where $g'_0(t)=\frac{dg_0(t)}{dt}$.
\begin{definition}
Given $g_0(t),h_0(t)\in \mathrm{S}(\mathbb{R}^+)$, we define the spectral functions $A(k)$ and $B(k)$ as follows:
\begin{eqnarray}
A(k)=\eta_2(0,k),~~B(k)=\eta_1(0,k), ~~k\in \overline{D}_1\cup \overline{D}_3,
\label{dAB}
\end{eqnarray}
where the vector function $\eta(t, k) = \left(\eta_1(t,k), \eta_2(t,k)\right)^T$ is the unique solution of
\begin{eqnarray}
\begin{split}
&\eta_{1,t}-\frac{i}{2}\left(k^{-1}+k\right)\eta_1=V_0^{11} \eta_1+V_0^{12} \eta_2,
\\
&\eta_{2,t}=V_0^{21} \eta_1-V_0^{11} \eta_2, ~~k\in \overline{D}_1\cup \overline{D}_3, ~~0<t<\infty,
\\
&\lim_{t\rightarrow \infty}\eta=(0,1)^T.
\end{split}
\label{dABeta}
\end{eqnarray}
Moreover, we define $\widetilde{A}(k)$ and $\widetilde{B}(k)$ in terms of $A(k)$ and $B(k)$ via the symmetries (\ref{symsfc}) and (\ref{symsfd}).
\end{definition}

\begin{proposition}
The spectral functions $A(k)$ and $B(k)$ have the following properties:
\\(1) A(k) and B(k) are analytic for $k\in D_1\cup D_3$, and continuous and bounded for $k\in \overline{D}_1\cup \overline{D}_3$.
\\(2) $A(k) = \exp\left(\frac{i}{4}\int_0^\infty\left(|g_0(\tau)|^2-|h_0(\tau)|^2\right)d\tau\right) + O(\frac{1}{k})$, $B(k) = O(\frac{1}{k})$, $k\rightarrow \infty$,
     $k\in \overline{D}_1\cup \overline{D}_3$.
\\(3) $A(k)\widetilde{A}(k)-B(k)\widetilde{B}(k)=1$, $k \in \left\{|k|=1\right\}\cup\mathbb{R}$.
\\(4) $(g_0(t),h_0(t))$ can be reconstructed in terms of $A(k)$ and $B(k)$ by
\begin{eqnarray}
\begin{split}
\bar{g}_0(t)\exp\left(\frac{i}{2}\int_t^\infty\left(|g_0(\tau)|^2-|h_0(\tau)|^2\right)d\tau\right)=\lim_{k\rightarrow \infty}(kM^{(t)}(t,k))^{12},
\\
-(2ig'_0(t)+g_0(t)|h_0(t)|^2+h_0(t))\exp\left(-\frac{i}{2}\int_t^\infty\left(|g_0(\tau)|^2-|h_0(\tau)|^2\right)d\tau\right)=\lim_{k\rightarrow \infty}(kM^{(t)}(t,k))^{21},
\end{split}
\label{ABis}
\end{eqnarray}
where $M^{(t)}(t, k)$ is the unique solution of the following RH problem:
\\$\bullet$
\begin{eqnarray}
M^{(t)}(t, k)=\left\{ \begin{array}{cc} M_{-}^{(t)}(t, k), ~~k\in \overline{D}_2\cup \overline{D}_4,   \\
  M_{+}^{(t)}(t, k), ~~k\in \overline{D}_1\cup \overline{D}_3,  \\ \end{array} \right.
\label{ABm}
\end{eqnarray}
is a sectionally meromorphic function.
\\$\bullet$
\begin{eqnarray}
M_{-}^{(t)}(t,k)=M_{+}^{(t)}(t,k)J^{(t)}(t,k), \quad k \in \left\{|k|=1\right\}\cup\mathbb{R},
\label{ABJc}
\end{eqnarray}
where
\begin{eqnarray}
J^{(t)}(t,k)=e^{\frac{i}{4}(k^{-1}+k)t\hat{\sigma_3}}\left( \begin{array}{cc} 1 & -\frac{B(k)}{\widetilde{A}(k)} \\
  \frac{\widetilde{B}(k)}{A(k)} & \frac{1}{A(k)\widetilde{A}(k)} \\ \end{array} \right).
\label{ABJr}
\end{eqnarray}
\\$\bullet$
\begin{eqnarray}
M^{(t)}(t,k)=I+O(\frac{1}{k}), \quad k\rightarrow \infty.
\label{ABMA}
\end{eqnarray}
\\$\bullet$
We assume that $A(k)$ can have at most $N$ simple zeros $\{\mathcal{K}_j\}_{1}^{N}$, where $\mathcal{K}_j\in D_{1}\cup D_{3}$;
$\widetilde{A}(k)$ can have at most $\widetilde{N}$ simple zeros $\{\widetilde{\mathcal{K}}_j\}_{1}^{\widetilde{N}}$,
 where $\widetilde{\mathcal{K}}_j\in D_{2}\cup D_{4}$.
The first column of $M_{+}^{(t)}(t,k)$ can have simple poles at $k=\mathcal{K}_j$, $j=1,\cdots,N$,
and the second column of $M_{-}^{(t)}(t,k)$ can have simple poles at $k=\widetilde{\mathcal{K}}_j$, $j=1,\cdots,\widetilde{N}$. The
associated residues are given by
\begin{subequations}
\begin{eqnarray}
{{\rm Res}}_{k=\mathcal{K}_j}[M^{(t)}(t, k)]^L=\frac{e^{-\frac{i}{2}(\mathcal{K}_j^{-1}+\mathcal{K}_j)t}}{\dot{A}(\mathcal{K}_j)B(\mathcal{K}_j)}[M^{(t)}(t, \mathcal{K}_j)]^R, ~~j=1,\cdots,N,
\label{ABrr1a}
\\
{{\rm Res}}_{k=\widetilde{\mathcal{K}}_j}[M^{(t)}(t,k)]^R=\frac{e^{\frac{i}{2}(\widetilde{\mathcal{K}}_j^{-1}+\widetilde{\mathcal{K}}_j)t}}
{\dot{\widetilde{A}}(\widetilde{\mathcal{K}}_j)\widetilde{B}(\widetilde{\mathcal{K}}_j)}
[M^{(t)}(t,\widetilde{\mathcal{K}}_j)]^{L},~~j=1,\cdots,\widetilde{N}.
\label{ABrr1b}
\end{eqnarray}
\label{ABrr1}
\end{subequations}
\end{proposition}
{\bf Proof}
Properties (1), (2) and (3) follow from definition 2.
The proof for property (4) is given in appendix C.  \QEDB

The main result for the problem in the complex $k$-plane is the following.
\begin{theorem}
Given $(u_0(x),v_0(x))\in \mathrm{S}(\mathbb{R}^+)$, define the spectral functions $a(k)$ and $b(k)$ by definition 1.
Suppose that there exists $(g_0(t),h_0(t))\in \mathrm{S}(\mathbb{R}^+)$ satisfying $u_0(0)=g_0(0)$, $v_0(0)=h_0(0)$,
such that the spectral functions $A(k)$ and $B(k)$, defined by definition 2,
satisfy the global relation (\ref{grk}). Assume that
\begin{itemize}
\item $a(k)$ has at most $n$ simple zeros $\{k_j\}_{1}^{n}$, $n=n_1+n_2$, where $k_j\in D_{1}$, $j=1,\cdots,n_1$; $k_j\in D_{2}$, $j=n_1+1,\cdots,n_1+n_2$.
 $\widetilde{a}(k)$ has at most $\widetilde{n}$ simple zeros $\{\widetilde{k}_j\}_{1}^{\widetilde{n}}$, $\widetilde{n}=\widetilde{n}_1+\widetilde{n}_2$, where $\widetilde{k}_j\in D_{4}$, $j=1,\cdots,\widetilde{n}_1$; $\widetilde{k}_j\in D_{3}$, $j=\widetilde{n}_1+1,\cdots,\widetilde{n}_1+\widetilde{n}_2$.
\item  $d(k)$ has at most $K$ simple zeros $\{\kappa_j\}_{1}^{K}$, where $\kappa_j\in D_{2}$, $j=1,\cdots,K$.
  $\widetilde{d}(k)$ has at most $\widetilde{K}$ simple zeros $\{\widetilde{\kappa}_j\}_{1}^{\widetilde{K}}$, where $\widetilde{\kappa}_j\in D_{3}$, $j=1,\cdots,\widetilde{K}$.
\item  None of the zeros of $a(k)$ for $k\in D_{2}$ coincides with any of the zeros of $d(k)$.
   None of the zeros of $\widetilde{a}(k)$ for $k\in D_{3}$ coincides with any of the zeros of $\widetilde{d}(k)$.
\end{itemize}
Define $M(x,t,k)$ as the solution of the following $2\times 2$ matrix RH problem:
\begin{itemize}
\item $M$ is sectionally meromorphic in $k\in \mathbb{C}\backslash L$, where $L=L_1\cup L_2\cup L_3\cup L_4$ is defined by (\ref{Jm1});
see figure 1 for the contour $L$.
\item $M$ satisfies the jump condition
\begin{eqnarray}
M_{-}(x,t,k)=M_{+}(x,t,k)J(x,t,k), \quad k\in L,
\label{RHPtheorem1}
\end{eqnarray}
where $M$ is $M_{-}$ for $k\in \overline{D}_2\cup \overline{D}_4$,
$M$ is $M_{+}$ for $k\in \overline{D}_1\cup \overline{D}_3$, and
$J$ is defined in terms of $\left\{a(k), b(k), A(k),B(k)\right\}$ by equations (\ref{Jm1}) and (\ref{JM1j});
see figure 1 for the domains $\{D_j\}_1^4$.
\item As $k\rightarrow \infty$,
\begin{eqnarray}
\begin{split}
M(x,t,k)\rightarrow I+O(\frac{1}{k}).
\end{split}
\label{Masptheorem1}
\end{eqnarray}
\item The residues associated with $M$ satisfy the relations in (\ref{rr1}).
\end{itemize}
Then $M(x, t, k)$ exists and is unique. Define $(u(x,t),v(x,t))$ in terms of $M(x, t, k)$ by
\begin{eqnarray}
\begin{split}
\bar{u}\exp\left(-\frac{i}{2}\int_x^\infty\left(|u(\xi,t)|^2+|v(\xi,t)|^2\right)d\xi\right)
=\lim_{k\rightarrow \infty}\left(k M^{12}(x,t,k)\right),
\\
\left(2iu_x+u|v|^2+v\right)\exp\left(\frac{i}{2}\int_x^\infty\left(|u(\xi,t)|^2+|v(\xi,t)|^2\right)d\xi\right)
=\lim_{k\rightarrow \infty}\left(kM^{21}(x,t,k)\right).
\end{split}
\label{uvstheorem1}
\end{eqnarray}
Then $(u(x,t),v(x,t))$ solves the MT equation (\ref{MT}) with the initial-boundary condition (\ref{IBV}).
\end{theorem}
{\bf Proof}
If $a(k)$, $d(k)$, $\widetilde{a}(k)$, $\widetilde{d}(k)$ have no zeros for $k\in D_1$, $k\in D_2$, $k\in D_3$, $k\in D_4$ respectively,
then the RH problem for $M(x,t,k)$ is regular. Its unique solvability is a consequence of the existence of a vanishing lemma; see \cite{Zhouxin2,FI,PS1}.
If $a(k)$, $d(k)$, $\widetilde{a}(k)$, $\widetilde{d}(k)$ can have a finite number of zeros,
this singular RH problem can be mapped to a regular one supplemented by a system of algebraic equations which is uniquely solvable; see \cite{FI}.

{\it Proof that $(u(x,t),v(x,t))$ satisfies the MT system.}
Based on the dressing method \cite{ZS}, it can be verified directly that if $M(x, t, k)$ is defined as the unique solution of the above RH problem,
and if $(u(x,t),v(x,t))$ is defined by the formula (\ref{uvstheorem1}), then $(u(x,t),v(x,t))$ satisfies the MT system; see appendix D.

{\it Proof that $u(x, 0)=u_0(x)$, $v(x, 0)=v_0(x)$.}
The proof that $(u(x,t),v(x,t))$ satisfies the initial condition follows from the fact that
it is possible to map the RH problem for $M(x, 0, k)$ to that for $M^{(x)}(x, k)$.
Let us define $M^{(x)}(x, k)$ by
\begin{subequations}
\begin{eqnarray}
M^{(x)}(x,k)=M(x,0,k), ~~k\in \overline{D}_1\cup \overline{D}_4,
\label{Mxtka}
\\
M^{(x)}(x,k)=M(x,0,k)J^{-1}_1(x,0,k), ~~k\in \overline{D}_2,
\label{Mxtkb}
\\
M^{(x)}(x,k)=M(x,0,k)J_3(x,0,k), ~~k\in \overline{D}_3.
\label{Mxtkc}
\end{eqnarray}
\label{Mxtk}
\end{subequations}
Then $M^{(x)}(x, k)$ satisfies the conditions (\ref{abJc}), (\ref{abJr}) and (\ref{abMA}).
We need to verify that $M^{(x)}(x, k)$ satisfies the residue conditions (\ref{abrr1}).
The first column of $M(x, t, k)$ has poles at $\{k_j\}_{1}^{n_1}$ for $k_j\in D_{1}$ and has poles at $\{\kappa_j\}_{1}^{K}$ for $\kappa_j\in D_{2}$;
while the first column of $M^{(x)}(x,k)$ should have poles at $\{k_j\}_{1}^{n}$, $n=n_1+n_2$, for $k_j\in D_{1}\cup D_{2}$.
We will now show that the transformations (\ref{Mxtk}) map the former poles to the
latter ones.
From (\ref{Mxtka}) we know $M^{(x)}(x,k)=M(x,0,k)$ for $k\in \overline{D}_1$, thus $M^{(x)}(x,k)$ has poles at $\{k_j\}_{1}^{n_1}$ with the
correct residue conditions (\ref{abrr1a}) for $j=1,\cdots, n_1$.
Equation (\ref{Mxtkb}) can be written as
\begin{eqnarray}
M^{(x)}(x,k)=\left(M^{L}(x,0,k)-e^{-\frac{i}{2}(k^{-1}-k)x}\Gamma(k)M^{R}(x,0,k),~M^{R}(x,0,k)\right).
\label{Mxtkb1}
\end{eqnarray}
Using this equation and the residue conditions (\ref{rr1c}), we find that $M^{(x)}(x,k)$ has no poles at $\{\kappa_j\}_{1}^{K}$.
Moreover, equation (\ref{Mxtkb1}) shows that $M^{(x)}(x,k)$ has poles at $\{k_j\}_{n_1+1}^{n}$ and the associated residues satisfy
\begin{eqnarray}
{{\rm Res}}_{k=k_j}[M^{(x)}(x, k)]^L=-{{\rm Res}}_{k=k_j}\Gamma(k)e^{-\frac{i}{2}(k_j^{-1}-k_j)x}[M^{(x)}(x, k_j)]^R, ~~j=n_1+1,\cdots,n.
\label{Mxtkb2}
\end{eqnarray}
Using the definitions of $\Gamma(k)$ and $d(k)$, equation (\ref{Mxtkb2}) becomes the residue conditions (\ref{abrr1a}) for $j=n_1+1,\cdots,n$.
In a similar manner, we can show that the transformations (\ref{Mxtk}) map the poles of the second column of $M(x, t, k)$
to the ones of $M^{(x)}(x,k)$.
Thus, $M^{(x)}(x, k)$ satisfies the same RH problem as that in proposition 1.
Comparing equation (\ref{abis}) with equation (\ref{uvstheorem1}) evaluated at $t=0$,
we obtain $u(x, 0)=u_0(x)$, $v(x, 0)=v_0(x)$.

{\it Proof that $u(0, t)=g_0(t)$, $v(0, t)=h_0(t)$.}
The proof that $(u(x,t),v(x,t))$ satisfies the boundary condition follows from the fact
that it is possible to map the RH problem for $M(0,t,k)$ to that for $M^{(t)}(t, k)$.
In order to show this, we introduce the transformation matrix
\begin{eqnarray}
\begin{split}
\Omega^{(1)}(t,k)=\left( \begin{array}{cc} \frac{a(k)}{A(k)} & 0 \\
 0 & \frac{A(k)}{a(k)} \\ \end{array} \right), ~~k\in \overline{D}_1,
 \\
\Omega^{(2)}(t,k)=\left( \begin{array}{cc} d(k) & -\frac{b(k)}{\widetilde{A}(k)}e^{\frac{i}{2}(k^{-1}+k)t} \\
 0 & \frac{1}{d(k)} \\ \end{array} \right), ~~k\in \overline{D}_2,
 \\
 \Omega^{(3)}(t,k)=\left( \begin{array}{cc} \frac{1}{\widetilde{d}(k)} & 0 \\
 -\frac{\widetilde{b}(k)}{A(k)}e^{-\frac{i}{2}(k^{-1}+k)t} & \widetilde{d}(k) \\ \end{array} \right), ~~k\in \overline{D}_3,
 \\
\Omega^{(4)}(t,k)=\left( \begin{array}{cc} \frac{\widetilde{A}(k)}{\widetilde{a}(k)} & 0 \\
 0 & \frac{\widetilde{a}(k)}{\widetilde{A}(k)} \\ \end{array} \right), ~~k\in \overline{D}_4,
\end{split}
\label{Omegaj}
\end{eqnarray}
and we define $M^{(t)}(t,k)$ by
\begin{eqnarray}
M^{(t)}(t,k)=F M(0,t,k)\Omega(t,k),
\label{Mttk}
\end{eqnarray}
where $\Omega(t,k)$ is $\Omega^{(j)}(t,k)$ for $k\in \overline{D}_{j}$, $j=1,\cdots,4$, and
\begin{eqnarray*}
F=\text{diag}
\left(
e^{\frac{i}{4}\left(\int_0^\infty\left(|u_0(x)|^2+|v_0(x)|^2\right)dx+\int_0^\infty\left(|g_0(t)|^2-|h_0(t)|^2\right)dt\right)},
e^{-\frac{i}{4}\left(\int_0^\infty\left(|u_0(x)|^2+|v_0(x)|^2\right)dx+\int_0^\infty\left(|g_0(t)|^2-|h_0(t)|^2\right)dt\right)}
\right).
\label{F}
\end{eqnarray*}
Using the global relation (\ref{grk}) we can verify directly that $\{\Omega^{(j)}(t,k)\}_1^4$ satisfy
\begin{eqnarray}
\begin{split}
&J_1(0,t,k)\Omega^{(2)}(t,k)=\Omega^{(1)}(t,k)J^{(t)}(t,k), ~~k\in L_1,
 \\
&J_2(0,t,k)\Omega^{(2)}(t,k)=\Omega^{(3)}(t,k)J^{(t)}(t,k), ~~k\in L_2,
 \\
 &J_3(0,t,k)\Omega^{(4)}(t,k)=\Omega^{(3)}(t,k)J^{(t)}(t,k), ~~k\in L_3,
 \\
&J_4(0,t,k)\Omega^{(4)}(t,k)=\Omega^{(1)}(t,k)J^{(t)}(t,k), ~~k\in L_4.
\end{split}
\label{OJr}
\end{eqnarray}
Equations (\ref{Mttk}) and (\ref{OJr}) imply that $M^{(t)}(t,k)$ satisfies exactly the jump condition (\ref{ABJc})
with the jump matrix defined by (\ref{ABJr}) and satisfies the estimate (\ref{ABMA}).
Moreover, in analogy with the proof used for $u(x, 0)=u_0(x)$, $v(x, 0)=v_0(x)$, one can verify that the transformation (\ref{Mttk})
replaces poles at $\{k_j\}_1^{n_1}$, $\{\kappa_j\}_{1}^{K}$, $\{\widetilde{k}_j\}_1^{\widetilde{n}_1}$,
$\{\widetilde{\kappa}_j\}_{1}^{\widetilde{K}}$
by poles at $\{\mathcal{K}_j\}_{1}^{N}$, $\{\widetilde{\mathcal{K}}_j\}_{1}^{\widetilde{N}}$,
with the residue conditions (\ref{rr1}) replaced by the residue conditions (\ref{ABrr1}).
Thus, $M^{(t)}(t, k)$ satisfies the same RH problem as that in the proposition 2.
\QEDB

\subsection{Inverse part for the problem in the complex $z$-plane}

The spectral analysis, performed for the Lax pair (\ref{LPz}) in the above section,
motivates the following definitions and properties for the spectral functions in the complex $z$-plane.

Let
\begin{eqnarray}
\begin{split}
\mathcal{U}_0(x,z)=&\left( \begin{array}{cc} \frac{i}{4}\left(|u_0(x)|^2+|v_0(x)|^2\right) & -\frac{i}{2}\bar{v}_0(x)\\
v'_0(x)+\frac{i}{2}|u_0(x)|^2v_0(x)+\frac{i}{2}u_0(x) &  -\frac{i}{4}\left(|u_0(x)|^2+|v_0(x)|^2\right) \\ \end{array} \right)
\\
&-\frac{i}{2z}\left( \begin{array}{cc}  \bar{u}_0(x)v_0(x) & -\bar{u}_0(x) \\
v_0(x)+\bar{u}_0(x)v^2_0(x) &  - \bar{u}_0(x)v_0(x) \\ \end{array} \right)
\\
\triangleq
& \left( \begin{array}{cc} \mathcal{U}_0^{11} & \mathcal{U}_0^{12} \\
\mathcal{U}_0^{21} &  -\mathcal{U}_0^{11} \\ \end{array} \right),
\end{split}
\label{U0z}
\end{eqnarray}
where $v'_0(x)=\frac{dv_0(x)}{dx}$.
\begin{definition}
Given $(u_0(x),v_0(x))\in \mathrm{S}(\mathbb{R}^+)$, we define the spectral functions $\mathfrak{a}(z)$ and $\mathfrak{b}(z)$ as follows:
\begin{eqnarray}
\mathfrak{a}(z)=\phi_2(0,z),~~\mathfrak{b}(z)=\phi_1(0,z), ~~{\rm Im} ~z\leq 0,
\label{dabz}
\end{eqnarray}
where the vector function $\mu(x, z) = \left(\mu_1(x,z), \mu_2(x,z)\right)^T$ is the unique solution of
\begin{eqnarray}
\begin{split}
&\mu_{1,x}-\frac{i}{2}\left(z-z^{-1}\right)\mu_1=\mathcal{U}_0^{11} \mu_1+\mathcal{U}_0^{12} \mu_2,
\\
&\mu_{2,x}=\mathcal{U}_0^{21} \mu_1-\mathcal{U}_0^{11} \mu_2, ~~{\rm Im} ~z\leq 0, ~~0<x<\infty,
\\
&\lim_{x\rightarrow \infty}\mu=(0,1)^T.
\end{split}
\label{dabmuz}
\end{eqnarray}
Moreover, we define $\widetilde{\mathfrak{a}}(z)$ and $\widetilde{\mathfrak{b}}(z)$ in terms of $\mathfrak{a}(z)$ and $\mathfrak{b}(z)$
via the symmetries (\ref{symsfaz}) and (\ref{symsfbz}).
\end{definition}

\begin{proposition}
The spectral functions $\mathfrak{a}(z)$ and $\mathfrak{b}(z)$ have the following properties:
\\(1) $\mathfrak{a}(z)$ and $\mathfrak{b}(z)$ are analytic for ${\rm Im} ~z<0$ and continuous and bounded for ${\rm Im} ~z\leq 0$.
\\(2) $\mathfrak{a}(z)=\exp\left(\frac{i}{4}\int_0^\infty\left(|u_0(\xi)|^2+|v_0(\xi)|^2\right)d\xi\right)+O(\frac{1}{z})$, $\mathfrak{b}(z)=O(\frac{1}{z})$, $z\rightarrow \infty$, ${\rm Im} ~z\leq 0$.
\\(3) $\mathfrak{a}(z)\widetilde{\mathfrak{a}}(z)-\mathfrak{\mathfrak{b}}(z)\widetilde{\mathfrak{b}}(z)=1$, $z \in \mathbb{R}$.
\\(4) $(u_0(x),v_0(x))$ can be reconstructed in terms of $\mathfrak{a}(z)$ and $\mathfrak{b}(z)$ by
\begin{eqnarray}
\begin{split}
\bar{v}_0(x)\exp\left(\frac{i}{2}\int_x^\infty\left(|u_0(\xi)|^2+|v_0(\xi)|^2\right)d\xi\right)=\lim_{z\rightarrow \infty}(z \mathcal{M}^{(x)}(x,z))^{12},
\\
\left(-2iv'_0(x)+|u_0(x)|^2v_0(x)+u_0(x)\right)\exp\left(-\frac{i}{2}\int_x^\infty\left(|u_0(\xi)|^2+|v_0(\xi)|^2\right)d\xi\right)=\lim_{z\rightarrow \infty}(z\mathcal{M}^{(x)}(x,z))^{21},
\end{split}
\label{abisz}
\end{eqnarray}
where $\mathcal{M}^{(x)}(x, z)$ is the unique solution of the following RH problem:
\\$\bullet$
\begin{eqnarray}
\mathcal{M}^{(x)}(x, z)=\left\{ \begin{array}{cc} \mathcal{M}_{-}^{(x)}(x, z), ~~{\rm Im} z\geq 0,   \\
  \mathcal{M}_{+}^{(x)}(x, z), ~~{\rm Im} z\leq 0,  \\ \end{array} \right.
\label{abmz}
\end{eqnarray}
is a sectionally meromorphic function.
\\$\bullet$
\begin{eqnarray}
\mathcal{M}_{-}^{(x)}(x,z)=\mathcal{M}_{+}^{(x)}(x,z)\mathcal{J}^{(x)}(x,z), \quad z\in \mathbb{R},
\label{abJcz}
\end{eqnarray}
where
\begin{eqnarray}
\mathcal{J}^{(x)}(x,z)=e^{\frac{i}{4}(z-z^{-1})x\hat{\sigma_3}}\left( \begin{array}{cc} 1 & -r(z) \\
  \widetilde{r}(z) & 1-r(z)\widetilde{r}(z) \\ \end{array} \right),
~r(z)=\frac{\mathfrak{b}(z)}{\widetilde{\mathfrak{a}}(z)}, ~\widetilde{r}(z)=\frac{\widetilde{\mathfrak{b}}(z)}{\mathfrak{a}(z)}.
\label{abJrz}
\end{eqnarray}
\\$\bullet$
\begin{eqnarray}
\mathcal{M}^{(x)}(x,z)=I+O(\frac{1}{z}), \quad z\rightarrow \infty.
\label{abMAz}
\end{eqnarray}
\\$\bullet$
We assume that $\mathfrak{a}(z)$ has at most $m$ simple zeros $\{z_j\}_{1}^{m}$, $m=m_1+m_2$, where $z_j\in \mathcal{D}_{1}$, $j=1,\cdots,m_1$; $z_j\in \mathcal{D}_{2}$, $j=m_1+1,\cdots,m_1+m_2$.
 $\widetilde{\mathfrak{a}}(z)$ has at most $\widetilde{m}$ simple zeros $\{\widetilde{z}_j\}_{1}^{\widetilde{m}}$, $\widetilde{m}=\widetilde{m}_1+\widetilde{m}_2$, where $\widetilde{z}_j\in \mathcal{D}_{4}$, $j=1,\cdots,\widetilde{m}_1$; $\widetilde{z}_j\in \mathcal{D}_{3}$, $j=\widetilde{m}_1+1,\cdots,\widetilde{m}_1+\widetilde{m}_2$.
The first column of $\mathcal{M}_{+}^{(x)}(x,z)$ can have simple poles at $z=z_j$, $j=1,\cdots,m$, and the second column of $\mathcal{M}_{-}^{(x)}(x,z)$
can have simple poles at $z=\widetilde{z}_j$, $j=1,\cdots,\widetilde{m}$. The
associated residues are given by
\begin{subequations}
\begin{eqnarray}
{{\rm Res}}_{z=z_j}[M^{(x)}(x, z)]^L=\frac{e^{-\frac{i}{2}(z_j-z_j^{-1})x}}{\dot{\mathfrak{a}}(z_j)\mathfrak{b}(z_j)}[\mathcal{M}^{(x)}(x, z_j)]^R, ~~j=1,\cdots,m,
\label{abrr1az}
\\
{{\rm Res}}_{z=\widetilde{z}_j}[M^{(x)}(x, z)]^R=\frac{e^{\frac{i}{2}(\widetilde{z}_j-\widetilde{z}_j^{-1})x}}{\dot{\widetilde{\mathfrak{a}}}(\widetilde{z}_j)\widetilde{\mathfrak{b}}(\widetilde{z}_j)}
[\mathcal{M}^{(x)}(x,\widetilde{z}_j)]^{L},~~j=1,\cdots,\widetilde{m}.
\label{abrr1bz}
\end{eqnarray}
\label{abrr1z}
\end{subequations}
\end{proposition}

Let
\begin{eqnarray}
\begin{split}
\mathcal{V}_0(t,z)=&\left( \begin{array}{cc} -\frac{i}{4}\left(|g_0(t)|^2-|h_0(t)|^2\right) & -\frac{i}{2}\bar{h}_0(t)\\
h'_0(t)-\frac{i}{2}|g_0(t)|^2h_0(t)-\frac{i}{2}g_0(t) &  \frac{i}{4}\left(|g_0(t)|^2-|h_0(t)|^2\right) \\ \end{array} \right)
\\
&+\frac{i}{2z}\left( \begin{array}{cc} \bar{g}_0(t) h_0(t) & -\bar{g}_0(t) \\
h_0(t)+\bar{g}_0(t)h^2_0(t) &  -\bar{g}_0(t) h_0(t) \\ \end{array} \right)
\\
\triangleq
& \left( \begin{array}{cc} \mathcal{V}_0^{11} & \mathcal{V}_0^{12} \\
\mathcal{V}_0^{21} &  -\mathcal{V}_0^{11} \\ \end{array} \right),
\end{split}
\label{V0z}
\end{eqnarray}
where $h'_0(t)=\frac{dh_0(t)}{dt}$.
\begin{definition}
Given $(g_0(t),h_0(t))\in \mathrm{S}(\mathbb{R}^+)$, we define the spectral functions $\mathcal{A}(z)$ and $\mathcal{B}(z)$ as follows:
\begin{eqnarray}
\mathcal{A}(z)=\chi_2(0,z),~~\mathcal{B}(z)=\chi_1(0,z), ~~z\in \overline{\mathcal{D}}_1\cup \overline{\mathcal{D}}_3,
\label{dABz}
\end{eqnarray}
where the vector function $\chi(t, z) = \left(\chi_1(t,z), \chi_2(t,z)\right)^T$ is the unique solution of
\begin{eqnarray}
\begin{split}
&\chi_{1,t}-\frac{i}{2}\left(z+z^{-1}\right)\chi_1=\mathcal{V}_0^{11} \chi_1+\mathcal{V}_0^{12} \chi_2,
\\
&\chi_{2,t}=\mathcal{V}_0^{21} \chi_1-\mathcal{V}_0^{11} \chi_2, ~~z\in \overline{\mathcal{D}}_1\cup \overline{\mathcal{D}}_3, ~~0<t<\infty,
\\
&\lim_{t\rightarrow \infty}\chi=(0,1)^T.
\end{split}
\label{dABchiz}
\end{eqnarray}
Moreover, we define $\widetilde{\mathcal{A}}(z)$ and $\widetilde{\mathcal{B}}(z)$ in terms of $\mathcal{A}(z)$ and $\mathcal{B}(z)$ via the symmetries (\ref{symsfcz}) and (\ref{symsfdz}).
\end{definition}

\begin{proposition}
The spectral functions $\mathcal{A}(z)$ and $\mathcal{B}(z)$ have the following properties:
\\(1) $\mathcal{A}(z)$ and $\mathcal{B}(z)$ are analytic for $z\in \mathcal{D}_1\cup \mathcal{D}_3$,
and continuous and bounded for $z\in \overline{\mathcal{D}}_1\cup \overline{\mathcal{D}}_3$.
\\(2) $\mathcal{A}(z)=\exp\left(-\frac{i}{4}\int_0^\infty\left(|g_0(\tau)|^2-|h_0(\tau)|^2\right)d\tau\right)+O(\frac{1}{z})$, $\mathcal{B}(z)=O(\frac{1}{z})$, $z\rightarrow \infty$, $z\in \overline{\mathcal{D}}_1\cup \overline{\mathcal{D}}_3$.
\\(3) $\mathcal{A}(z)\widetilde{\mathcal{A}}(z)-\mathcal{B}(z)\widetilde{\mathcal{B}}(z)=1$, $z \in \left\{|z|=1\right\}\cup\mathbb{R}$.
\\(4) $(g_0(t),h_0(t))$ can be reconstructed in terms of $\mathcal{A}(z)$ and $\mathcal{B}(z)$ by
\begin{eqnarray}
\begin{split}
\bar{h}_0(t)\exp\left(-\frac{i}{2}\int_t^\infty\left(|g_0(\tau)|^2-|h_0(\tau)|^2\right)d\tau\right)=\lim_{z\rightarrow \infty}(z\mathcal{M}^{(t)}(t,z))^{12},
\\
-(2ih'_0(t)+h_0(t)|g_0(t)|^2+g_0(t))\exp\left(\frac{i}{2}\int_t^\infty\left(|g_0(\tau)|^2-|h_0(\tau)|^2\right)d\tau\right)=\lim_{z\rightarrow \infty}(z\mathcal{M}^{(t)}(t,z))^{21},
\end{split}
\label{ABisz}
\end{eqnarray}
where $\mathcal{M}^{(t)}(t, z)$ is the unique solution of the following RH problem:
\\$\bullet$
\begin{eqnarray}
\mathcal{M}^{(t)}(t, z)=\left\{ \begin{array}{cc} \mathcal{M}_{-}^{(t)}(t, z), ~~z\in \overline{\mathcal{D}}_2\cup \overline{\mathcal{D}}_4,   \\
  \mathcal{M}_{+}^{(t)}(t, z), ~~z\in \overline{\mathcal{D}}_1\cup \overline{\mathcal{D}}_3,  \\ \end{array} \right.
\label{ABmz}
\end{eqnarray}
is a sectionally meromorphic function.
\\$\bullet$
\begin{eqnarray}
\mathcal{M}_{-}^{(t)}(t,z)=\mathcal{M}_{+}^{(t)}(t,z)\mathcal{J}^{(t)}(t,z), \quad z \in \left\{|z|=1\right\}\cup\mathbb{R},
\label{ABJcz}
\end{eqnarray}
where
\begin{eqnarray}
\mathcal{J}^{(t)}(t,z)=e^{\frac{i}{4}(z+z^{-1})t\hat{\sigma_3}}\left( \begin{array}{cc} 1 & -\frac{\mathcal{B}(z)}{\widetilde{\mathcal{A}}(z)} \\
  \frac{\widetilde{\mathcal{B}}(z)}{\mathcal{A}(z)} & \frac{1}{\mathcal{A}(z)\widetilde{\mathcal{A}}(z)} \\ \end{array} \right).
\label{ABJrz}
\end{eqnarray}
\\$\bullet$
\begin{eqnarray}
\mathcal{M}^{(t)}(t,z)=I+O(\frac{1}{z}), \quad z\rightarrow \infty.
\label{ABMAz}
\end{eqnarray}
\\$\bullet$
We assume that $\mathcal{A}(z)$ can have at most $\mathcal{N}$ simple zeros $\{\mathcal{Z}_j\}_{1}^{\mathcal{N}}$, where $\mathcal{Z}_j\in \mathcal{D}_{1}\cup \mathcal{D}_{3}$;
$\widetilde{\mathcal{A}}(z)$ can have at most $\widetilde{\mathcal{N}}$ simple zeros $\{\widetilde{\mathcal{Z}}_j\}_{1}^{\widetilde{\mathcal{N}}}$,
 where $\widetilde{\mathcal{Z}}_j\in \mathcal{D}_{2}\cup \mathcal{D}_{4}$.
The first column of $\mathcal{M}_{+}^{(t)}(t,z)$ can have simple poles at $z=\mathcal{Z}_j$, $j=1,\cdots,\mathcal{N}$,
and the second column of $\mathcal{M}_{-}^{(t)}(t,z)$ can have simple poles at $z=\widetilde{\mathcal{Z}}_j$, $j=1,\cdots,\widetilde{\mathcal{N}}$. The
associated residues are given by
\begin{subequations}
\begin{eqnarray}
{{\rm Res}}_{z=\mathcal{Z}_j}[\mathcal{M}^{(t)}(t, z)]^L=\frac{e^{-\frac{i}{2}(\mathcal{Z}_j+\mathcal{Z}_j^{-1})t}}{\dot{\mathcal{A}}(\mathcal{Z}_j)\mathcal{B}(\mathcal{Z}_j)}[\mathcal{M}^{(t)}(t, \mathcal{Z}_j)]^R, ~~j=1,\cdots,\mathcal{N},
\label{ABrr1az}
\\
{{\rm Res}}_{z=\widetilde{\mathcal{Z}}_j}[\mathcal{M}^{(t)}(t,z)]^R=\frac{e^{\frac{i}{2}(\widetilde{\mathcal{Z}}_j+\widetilde{\mathcal{Z}}_j^{-1})t}}
{\dot{\widetilde{\mathcal{A}}}(\widetilde{\mathcal{Z}}_j)\widetilde{\mathcal{B}}(\widetilde{\mathcal{Z}}_j)}
[\mathcal{M}^{(t)}(t,\widetilde{\mathcal{Z}}_j)]^{L},~~j=1,\cdots,\widetilde{\mathcal{N}}.
\label{ABrr1bz}
\end{eqnarray}
\label{ABrr1z}
\end{subequations}
\end{proposition}

The main result for the problem in the complex $z$-plane is the following.
\begin{theorem}
Given $(u_0(x),v_0(x))\in \mathrm{S}(\mathbb{R}^+)$, define the spectral functions $\mathfrak{a}(z)$ and $\mathfrak{b}(z)$ by definition 3.
Suppose that there exist $(g_0(t),h_0(t))\in \mathrm{S}(\mathbb{R}^+)$ satisfying $u_0(0)=g_0(0)$ and $v_0(0)=h_0(0)$,
such that the spectral functions $\mathcal{A}(z)$ and $\mathcal{B}(z)$, defined by definition 4,
satisfy the global relation (\ref{grz}). Assume that
\begin{itemize}
\item $\mathfrak{a}(z)$ has at most $m$ simple zeros $\{z_j\}_{1}^{m}$, $m=m_1+m_2$, where $z_j\in \mathcal{D}_{1}$, $j=1,\cdots,m_1$; $z_j\in \mathcal{D}_{2}$, $j=m_1+1,\cdots,m_1+m_2$.
 $\widetilde{\mathfrak{a}}(z)$ has at most $\widetilde{m}$ simple zeros $\{\widetilde{z}_j\}_{1}^{\widetilde{m}}$, $\widetilde{m}=\widetilde{m}_1+\widetilde{m}_2$, where $\widetilde{z}_j\in \mathcal{D}_{4}$, $j=1,\cdots,\widetilde{m}_1$; $\widetilde{z}_j\in \mathcal{D}_{3}$, $j=\widetilde{m}_1+1,\cdots,\widetilde{m}_1+\widetilde{m}_2$.
\item  $\alpha(z)$ has at most $\Lambda$ simple zeros $\{\nu_j\}_{1}^{\Lambda}$, where $\nu_j\in \mathcal{D}_{2}$, $j=1,\cdots,\Lambda$.
  $\widetilde{\alpha}(z)$ has at most $\widetilde{\Lambda}$ simple zeros $\{\widetilde{\nu}_j\}_{1}^{\widetilde{\Lambda}}$, where $\widetilde{\nu}_j\in \mathcal{D}_{3}$, $j=1,\cdots,\widetilde{\Lambda}$.
\item  None of the zeros of $\mathfrak{a}(z)$ for $z\in \mathcal{D}_{2}$ coincides with any of the zeros of $\alpha(z)$.
   None of the zeros of $\widetilde{\mathfrak{a}}(z)$ for $z\in \mathcal{D}_{3}$ coincides with any of the zeros of $\widetilde{\alpha}(z)$.
\end{itemize}
Define $\mathcal{M}(x,t,z)$ as the solution of the following $2\times 2$ matrix RH problem:
\begin{itemize}
\item $\mathcal{M}$ is sectionally meromorphic in $z\in \mathbb{C}\backslash \mathcal{L}$, where $\mathcal{L}=\mathcal{L}_1\cup \mathcal{L}_2\cup \mathcal{L}_3\cup \mathcal{L}_4$ is defined by (\ref{Jm2});
see figure 2 for the contour $\mathcal{L}$.
\item $\mathcal{M}$ satisfies the jump condition
\begin{eqnarray}
\mathcal{M}_{-}(x,t,z)=\mathcal{M}_{+}(x,t,z)\mathcal{J}(x,t,z), \quad z\in \mathcal{L},
\label{RHPtheorem1z}
\end{eqnarray}
where $\mathcal{M}$ is $\mathcal{M}_{-}$ for $z\in \overline{\mathcal{D}}_2\cup \overline{\mathcal{D}}_4$,
$\mathcal{M}$ is $\mathcal{M}_{+}$ for $z\in \overline{\mathcal{D}}_1\cup \overline{\mathcal{D}}_3$, and
$\mathcal{J}$ is defined in terms of $\left\{\mathfrak{a}(z), \mathfrak{b}(z), \mathcal{A}(z),\mathcal{B}(z)\right\}$ by equations (\ref{Jm2}) and (\ref{JM2j});
see figure 2 for the domains $\{\mathcal{D}_j\}_1^4$.
\item As $z\rightarrow \infty$,
\begin{eqnarray}
\begin{split}
\mathcal{M}(x,t,z)\rightarrow I+O(\frac{1}{z}).
\end{split}
\label{Masptheorem1z}
\end{eqnarray}
\item The residues associated with $\mathcal{M}$ satisfy the relations in (\ref{rr2}).
\end{itemize}
Then $\mathcal{M}(x, t, z)$ exists and is unique. Define $(u(x,t),v(x,t))$ in terms of $\mathcal{M}(x, t, z)$ by
\begin{eqnarray}
\begin{split}
\bar{v}\exp\left(\frac{i}{2}\int_x^\infty\left(|u(\xi,t)|^2+|v(\xi,t)|^2\right)d\xi\right)=\lim_{z\rightarrow \infty}\left(z\mathcal{M}^{12}(x,t,z)\right),
\\
\left(-2iv_x+|u|^2v+u\right)\exp\left(-\frac{i}{2}\int_x^\infty\left(|u(\xi,t)|^2+|v(\xi,t)|^2\right)d\xi\right)=\lim_{z\rightarrow \infty}\left(z\mathcal{M}^{21}(x,t,z)\right).
\end{split}
\label{uvstheorem1z}
\end{eqnarray}
Then $(u(x,t),v(x,t))$ solves the MT equation (\ref{MT}) with initial-boundary condition (\ref{IBV}).
\end{theorem}
{\bf Proof}
The proof for this theorem is similar to that of theorem 1.
\QEDB

\section {Conclusions}

By implementing the UTM,
we have expressed the solution of the IBV problem (\ref{IBV}) for the MT system (\ref{MT})
in terms of solutions of appropriate RH problems which have explicit $(x, t)$-dependence
and are defined in terms of the given initial and boundary values.
Our results show that, for the MT system,
the UTM for analyzing IBV problem is as effective as the IST for analyzing initial value problem.

We end this paper with the following remarks.

1. It is clear that the method can be generalized to analyze the IBV problem for the MT system on the finite interval,
namely, the MT system posed in the domain
$$\left\{(x,t)\in \mathbb{R}^2\mid~0\leq x\leq L,0\leq t<\infty\right\}.$$

2. By using the nonlinear steepest descent method \cite{DZ1},
long-time asymptotics of the solution of the initial value problem of the MT system has been derived recently in \cite{Saa}
with aid of the RH problems established in \cite{PS1}.
We note that the RH problems established in this paper for the IBV problem of the MT system have explicit exponential $(x, t)$-dependence.
Thus it is possible to study long-time asymptotics of the solution of the IBV problem of the MT system with aid of our RH problems. 
This issue is beyond the scope of the present paper, it is left for future study. 

3. We note that integrable nonlocal nonlinear equations were introduced recently by Ablowitz and Musslimani in \cite{AM1,AM2,AM3}.
Here we point out that the MT system (\ref{MT}) admits the nonlocal reduction $v(x,t)=\pm u(-x,t)$.
Using this reduction we find the following new nonlocal equation
\begin{eqnarray}
i\left(u_t(x,t)+u_x(x,t)\right) \pm u(-x,t)+|u(-x,t)|^2u(x,t)=0.
\label{nlMT}
\end{eqnarray}
The IST for this nonlocal MT equation is left for future investigation.


\section*{ACKNOWLEDGMENTS}

This work was supported by the National Natural Science Foundation of China (Grant No. 11771186).

\begin{appendices}
\section{Asymptotic behavior of the eigenfunctions $\{\Phi_j(x,t,k)\}_1^3$}

We consider the expansions of $\{\Phi_j(x,t,k)\}_1^3$ in the following form
\begin{eqnarray}
\Phi_j(x,t,k)=\Phi_j^{(0)}(x,t)+\Phi_j^{(1)}(x,t)\frac{1}{k}+\Phi_j^{(2)}(x,t)\frac{1}{k^2}+\cdots, ~~j=1,2,3.
\label{A1}
\end{eqnarray}
Substituting the above expansions into  (\ref{LPk}) and matching the powers of $k$, we find
\begin{subequations}
\begin{eqnarray}
\left[\sigma_3,\Phi_j^{(0)}\right]=0, ~~j=1,2,3,
\label{A2a}
\\
\Phi_{j,x}^{(0)}+\frac{i}{4}\left[\sigma_3,\Phi_j^{(1)}\right]=U_1\Phi_j^{(0)}, ~~j=1,2,3,
\label{A2b}
\\
\Phi_{j,t}^{(0)}-\frac{i}{4}\left[\sigma_3,\Phi_j^{(1)}\right]=V_1\Phi_j^{(0)}, ~~j=1,2,3.
\label{A2c}
\end{eqnarray}
\label{A2}
\end{subequations}
From (\ref{A2a}), we obtain
\begin{eqnarray}
\left(\Phi_j^{(0)}\right)^{12}=\left(\Phi_j^{(0)}\right)^{21}=0, ~~j=1,2,3,
\label{A2as}
\end{eqnarray}
where
$\left(\Phi_j^{(0)}\right)^{kl}$, $k,l=1,2$, denotes the $(k,l)$-entry of the $2\times 2$ matrix $\Phi_j^{(0)}$.
From (\ref{A2b}) and (\ref{A2c}), we find that, for $j=1,2,3$,
\begin{subequations}
\begin{eqnarray}
\left(\Phi_{j,x}^{(0)}\right)^{11}=-\frac{i}{4}\left(|u|^2+|v|^2\right)\left(\Phi_j^{(0)}\right)^{11},
~~\left(\Phi_{j,t}^{(0)}\right)^{11}=\frac{i}{4}\left(|u|^2-|v|^2\right)\left(\Phi_j^{(0)}\right)^{11},
\label{A2bcsa}
\\
\left(\Phi_{j,x}^{(0)}\right)^{22}=\frac{i}{4}\left(|u|^2+|v|^2\right)\left(\Phi_j^{(0)}\right)^{22},
~~\left(\Phi_{j,t}^{(0)}\right)^{22}=-\frac{i}{4}\left(|u|^2-|v|^2\right)\left(\Phi_j^{(0)}\right)^{22},
\label{A2bcsb}
\\
\left(\Phi_j^{(1)}\right)^{12}=\bar{u}\left(\Phi_j^{(0)}\right)^{22},
\label{A2bcsc}
\\
\left(\Phi_j^{(1)}\right)^{21}=2i\left(u_x-\frac{i}{2}u|v|^2-\frac{i}{2}v\right)\left(\Phi_j^{(0)}\right)^{11}
=-2i\left(u_t-\frac{i}{2}u|v|^2-\frac{i}{2}v\right)\left(\Phi_j^{(0)}\right)^{11},
\label{A2bcsd}
\end{eqnarray}
\label{A2bcs}
\end{subequations}
Using (\ref{A2bcsa}) and the normalization condition of $\{\Phi_j(x,t,k)\}_1^3$,
we obtain (\ref{aspPhi1a}), (\ref{aspPhi1b}) and (\ref{aspPhi1c}).
Using (\ref{A2bcsb}), (\ref{A2bcsc}) and (\ref{A2bcsd}), we find (\ref{aspPhi1d}), (\ref{aspPhi1e}) and (\ref{aspPhi1f}), respectively.

\section{ Proof of property (4) in Proposition 1}
We now derive property (4) in Proposition 1.
We introduce the vector function $\widetilde{\phi}(x,k)$ by
\begin{eqnarray}
\widetilde{\phi}(x,k)=\left( \begin{array}{cc} -\bar{u}_0(x) & 1 \\
-k-|u_0(x)|^2 &  u_0(x) \\ \end{array} \right)
\left( \begin{array}{c} \overline{\phi_1(x,\bar{k})} \\
\overline{\phi_2(x,\bar{k})} \\ \end{array} \right),
\label{dabphitilde}
\end{eqnarray}
and we let
\begin{eqnarray}
\Phi_3(x,k)=\left(\widetilde{\phi}(x,k), ~ \phi(x,k)\right).
\label{abproofPhi3}
\end{eqnarray}
Moreover, we define the matrix function $\Phi_2(x,k)=\left(\psi(x,k), ~ \widetilde{\psi}(x,k)\right)$ as the unique solution of
\begin{eqnarray}
\begin{split}
\Phi_{2,x}-\frac{i}{4}(k^{-1}-k)[\sigma_3,\Phi_2]=
U_0(x,k)\Phi_2,
\\
\Phi_2(0,k)=I_2,
\end{split}
\label{Psi2xk}
\end{eqnarray}
where $U_0(x,k)$ are defined by (\ref{U0}).
The matrix functions $\Phi_2(x,k)$ and $\Phi_3(x,k)$ satisfy the same matrix equation
$\Phi_{x}-\frac{i}{4}(k^{-1}-k)[\sigma_3,\Phi]=
U_0(x,k)\Phi$,
thus they are related:
\begin{eqnarray}
\Phi_{3}(x,k)=\Phi_{2}(x,k)e^{\frac{i}{4}(k^{-1}-k)x\hat{\sigma_3}}s(k), ~~k\in \mathbb{R}.
\label{sproof}
\end{eqnarray}
Let
\begin{eqnarray}
\begin{split}
 M_{-}^{(x)}(x, k)=E(x)\left(\widetilde{\phi}(x,k), ~ \frac{\widetilde{\psi}(x,k)}{\widetilde{a}(k)}\right), ~~{\rm Im} k\leq 0,
 \\
 M_{+}^{(x)}(x, k)=E(x)\left(\frac{\psi(x,k)}{a(k)}, ~\phi(x,k)\right), ~~{\rm Im} k\geq 0,
\end{split}
\label{abmproof}
\end{eqnarray}
where
\begin{eqnarray}
E(x)=\text{diag}\left(\exp\left(-\frac{i}{4}\int_x^\infty\left(|u_0(\xi)|^2+|v_0(\xi)|^2\right)d\xi\right),
\exp\left(\frac{i}{4}\int_x^\infty\left(|u_0(\xi)|^2+|v_0(\xi)|^2\right)d\xi\right)\right).
\label{Ex}
\end{eqnarray}
Equation (\ref{sproof}) can be rewritten as the jump condition (\ref{abJc}) with the jump matrix $J^{(x)}(x,k)$ defined by (\ref{abJr}).
We need to verify the residue conditions (\ref{abrr1}).
The second column of (\ref{sproof}) implies the following equation
\begin{eqnarray}
\phi(x,k)=a(k)\widetilde{\psi}(x,k)+e^{\frac{i}{2}(k^{-1}-k)x}b(k)\psi(x,k).
\label{phipsi}
\end{eqnarray}
The functions $\psi(x,k)$ and $\widetilde{\psi}(x,k)$ are analytic for $k\in \mathbb{C}\setminus\{0\}$.
Thus we can evaluate the above equation at $k=k_j$, this yields
\begin{eqnarray}
\phi(x,k_j)=e^{\frac{i}{2}(k_j^{-1}-k_j)x}b(k_j)\psi(x,k_j).
\label{phipsikj}
\end{eqnarray}
From the definition (\ref{abmproof}) and equation (\ref{phipsikj}), we find the residue conditions (\ref{abrr1a}).
The residue conditions (\ref{abrr1b}) can be derived similarly.

Moreover, (\ref{abmproof}) implies the following asymptotic expansion
\begin{eqnarray}
\begin{split}
M^{(x)}(x,k)= I+M^{(x)}_1(x)\frac{1}{k}+O(\frac{1}{k^2}),  ~~ k\rightarrow \infty,
\end{split}
\label{Masp1x}
\end{eqnarray}
where the $(1,2)$-entry and $(2,1)$-entry of the matrix $M^{(x)}_1(x)$ are given by
\begin{eqnarray}
\begin{split}
\left(M^{(x)}_1(x)\right)^{12}=\bar{u}_0(x)\exp\left(-\frac{i}{2}\int_x^\infty\left(|u_0(\xi)|^2+|v_0(\xi)|^2\right)d\xi\right),
\\
\left(M^{(x)}_1(x)\right)^{21}=(2iu'_0(x)+u_0(x)|v_0(x)|^2+v_0(x))\exp\left(\frac{i}{2}\int_x^\infty\left(|u_0(\xi)|^2+|v_0(\xi)|^2\right)d\xi\right).
\end{split}
\label{M112x}
\end{eqnarray}
The reconstruction formula (\ref{abis}) for $(u_0(x),v_0(x))$ follows from (\ref{Masp1x}) and (\ref{M112x}).
\QEDB

\section{ Proof of property (4) in Proposition 2}

We now derive property (4) in Proposition 2.
We introduce the vector function $\widetilde{\eta}(t,k)$ by
\begin{eqnarray}
\widetilde{\eta}(t,k)=\left( \begin{array}{cc} -\bar{g}_0(t) & 1 \\
-k-|g_0(t)|^2 &  g_0(t) \\ \end{array} \right)
\left( \begin{array}{c} \overline{\eta_1(t,\bar{k})} \\
\overline{\eta_2(t,\bar{k})} \\ \end{array} \right),
\label{dABetatilde}
\end{eqnarray}
and we let
\begin{eqnarray}
\Phi_1(t,k)=\left(\widetilde{\eta}(t,k), ~ \eta(t,k)\right).
\label{ABproofPhi3}
\end{eqnarray}
Moreover, we define the matrix function $\Phi_2(t,k)=\left(\rho(t,k), ~ \widetilde{\rho}(t,k)\right)$ as the unique solution of
\begin{eqnarray}
\begin{split}
\Phi_{2,t}-\frac{i}{4}(k^{-1}+k)[\sigma_3,\Phi_2]=
V_0(t,k)\Phi_2,
\\
\Phi_2(0,k)=I_2,
\end{split}
\label{Psi2tk}
\end{eqnarray}
where $V_0(t,k)$ are defined by (\ref{V0}).
The matrix functions $\Phi_1(t,k)$ and $\Phi_2(t,k)$ satisfy the same matrix equation
$\Phi_{t}-\frac{i}{4}(k^{-1}+k)[\sigma_3,\Phi]=
V_0(t,k)\Phi$,
thus they are related:
\begin{eqnarray}
\Phi_{1}(t,k)=\Phi_{2}(t,k)e^{\frac{i}{4}(k^{-1}+k)t\hat{\sigma_3}}S(k), ~~k\in \left\{|k|=1\right\}\cup\mathbb{R}.
\label{Sproof}
\end{eqnarray}
Let
\begin{eqnarray}
\begin{split}
 M_{-}^{(t)}(t, k)=E(t)\left(\widetilde{\eta}(t,k), ~ \frac{\widetilde{\rho}(t,k)}{\widetilde{A}(k)}\right), ~~k\in \overline{D}_2\cup \overline{D}_4,
 \\
 M_{+}^{(t)}(t, k)=E(t)\left(\frac{\rho(t,k)}{A(k)}, ~\eta(t,k)\right), ~~k\in \overline{D}_1\cup \overline{D}_3,
\end{split}
\label{ABmproof}
\end{eqnarray}
where
\begin{eqnarray}
E(t)=\text{diag}\left(\exp\left(\frac{i}{4}\int_t^\infty\left(|g_0(\tau)|^2-|h_0(\tau)|^2\right)d\tau\right),
\exp\left(-\frac{i}{4}\int_t^\infty\left(|g_0(\tau)|^2-|h_0(\tau)|^2\right)d\tau\right)\right).
\label{Et}
\end{eqnarray}
Equation (\ref{Sproof}) can be rewritten as the jump condition (\ref{ABJc}) with the jump matrix $J^{(t)}(t,k)$ defined by (\ref{ABJr}).
The proof for the residue conditions (\ref{ABrr1}) is the same as in the case of the function $M^{(x)}(x, k)$.

Moreover, from (\ref{ABmproof}) we find the following asymptotic expansion
\begin{eqnarray}
\begin{split}
M^{(t)}(t,k)= I+M^{(t)}_1(t)\frac{1}{k}+O(\frac{1}{k^2}),  ~~ k\rightarrow \infty,
\end{split}
\label{Masp1t}
\end{eqnarray}
where the $(1,2)$-entry and $(2,1)$-entry of the matrix $M^{(x)}_1(x)$ are given by
\begin{eqnarray}
\begin{split}
\left(M^{(t)}_1(t)\right)^{12}=\bar{g}_0(t)\exp\left(\frac{i}{2}\int_t^\infty\left(|g_0(\tau)|^2-|h_0(\tau)|^2\right)d\tau\right),
\\
\left(M^{(t)}_1(t)\right)^{21}=-(2ig'_0(t)+g_0(t)|h_0(t)|^2+h_0(t))\exp\left(-\frac{i}{2}\int_t^\infty\left(|g_0(\tau)|^2-|h_0(\tau)|^2\right)d\tau\right).
\end{split}
\label{M112t}
\end{eqnarray}
Using (\ref{Masp1t}) and (\ref{M112t}) we find the reconstruction formula (\ref{ABis}) for $(g_0(t),h_0(t))$.

\section{Proof that (u,v) defined by (\ref{uvstheorem1}) satisfies the MT system}
We introduce
\begin{eqnarray}
q(x,t)=\lim_{k\rightarrow \infty}\left(k M^{12}(x,t,k)\right), ~~p(x,t)=\lim_{k\rightarrow \infty}\left(kM^{21}(x,t,k)\right).
\label{qp}
\end{eqnarray}
We define a pair of linear operators $\mathfrak{L}_1$ and $\mathfrak{L}_2$ by
\begin{eqnarray}
\begin{split}
\mathfrak{L}_1M=\partial_xM-\frac{i}{4}(k^{-1}-k)[\sigma_3,M]-\frac{i}{2}Q_1M-\frac{i}{2k}Q_2M,
\\
\mathfrak{L}_2M=\partial_tM-\frac{i}{4}(k^{-1}+k)[\sigma_3,M]-\frac{i}{2}P_1M-\frac{i}{2k}P_2M,
\end{split}
\label{LPA}
\end{eqnarray}
where
\begin{eqnarray}
\begin{split}
Q_1=\left( \begin{array}{cc} 0 & q \\
 -p &  0 \\ \end{array} \right),
~~P_1=-Q_1,
\\
Q_2=\left( \begin{array}{cc} \bar{q}(\bar{p}+2iq_x+q|q|^2) & -(\bar{p}+2iq_x+q|q|^2) \\
 \bar{q}+(\bar{q})^2(\bar{p}+2iq_x+q|q|^2) &  -\bar{q}(\bar{p}+2iq_x+q|q|^2) \\ \end{array} \right),
\\
 P_2=\left( \begin{array}{cc} \bar{q}(2iq_t-\bar{p}-q|q|^2) & -(2iq_t-\bar{p}-q|q|^2) \\
 \bar{q}+(\bar{q})^2(2iq_t-\bar{p}-q|q|^2) &  -\bar{q}(2iq_t-\bar{p}-q|q|^2) \\ \end{array} \right).
\end{split}
\label{QP12}
\end{eqnarray}
It can be verified directly that $\mathfrak{L}_1M$ and $\mathfrak{L}_2M$ satisfy  the same jump condition as $M$ defined in theorem 1,  i.e.,
\begin{eqnarray}
\begin{split}
\mathfrak{L}_1M_{-}(x,t,k)=\left(\mathfrak{L}_1M_{+}(x,t,k)\right)J(x,t,k),
\\
\mathfrak{L}_2M_{-}(x,t,k)=\left(\mathfrak{L}_2M_{+}(x,t,k)\right)J(x,t,k).
\end{split}
\label{RHP1AppJ}
\end{eqnarray}
Moreover, as $k\rightarrow \infty$,
\begin{eqnarray}
\mathfrak{L}_1M(x,t,k)=O(\frac{1}{k}),
~~
\mathfrak{L}_2M(x,t,k)=O(\frac{1}{k}).
\label{RHP1AppAS}
\end{eqnarray}
The uniqueness of the solution of this RH problem implies that $\mathfrak{L}_1M=0$ and $\mathfrak{L}_2M=0$, namely,
\begin{eqnarray}
\begin{split}
\partial_xM-\frac{i}{4}(k^{-1}-k)[\sigma_3,M]-\frac{i}{2}Q_1M-\frac{i}{2k}Q_2M=0,
\\
\partial_tM-\frac{i}{4}(k^{-1}+k)[\sigma_3,M]-\frac{i}{2}P_1M-\frac{i}{2k}P_2M=0.
\end{split}
\label{LPA2}
\end{eqnarray}
The compatibility condition of (\ref{LPA2}) yields the following equation
\begin{eqnarray}
\begin{split}
q_t-q_x+i\bar{p}+iq|q|^2=0,
\\
p_t+p_x-i\bar{q}-i(\bar{q})^2(\bar{p}+2iq_x+q|q|^2)=0.
\end{split}
\label{NE1}
\end{eqnarray}
After straightforward calculations, we find that the transformation
\begin{eqnarray}
\begin{split}
q=\bar{u}\exp\left(-\frac{i}{2}\int_x^\infty\left(|u(\xi,t)|^2+|v(\xi,t)|^2\right)d\xi\right),
\\
p=\left(2iu_x+u|v|^2+v\right)\exp\left(\frac{i}{2}\int_x^\infty\left(|u(\xi,t)|^2+|v(\xi,t)|^2\right)d\xi\right),
\end{split}
\label{uvpq}
\end{eqnarray}
maps the equation (\ref{NE1}) to the MT system (\ref{MT}).

\end{appendices}

\vspace{1cm}
\small{

}
\end{document}